\documentclass[]{aa}  
\usepackage{graphicx}
\usepackage{txfonts}

\def\source {1RXS J170849.0$-$400910}
\def\Bp {{B_{\rm p}}}
\def\Rns {{R_{\rm NS}}}
\def\der {{\rm d}}
\def\Dfns {\Delta\phi_{\rm N-S}}

\def\me {m_{\rm e}}
\def\Bq {B_{\rm q}}

\begin{document}

   \title{Fitting {\it XMM-Newton} Observations of the AXP 1RXS J170849.0$-$400910 with four magnetar surface emission models, and predictions for X-ray polarization observations with {\it IXPE}}
   \titlerunning{Fitting magnetar observations with four surface emission models}

   \author{Henric\,Krawczynski\inst{1},
          Roberto Taverna\inst{2,3},
          Roberto Turolla\inst{3,4},
          Sandro Mereghetti\inst{5}, \and
          Michela Rigoselli\inst{5}
          }
   \institute{Physics Department and McDonnell Center for the Space Sciences, Washington University in St. Louis, MO, 63130, USA
         \and
             Department of Mathematics and Physics, University of Roma Tre, via della Vasca Navale 84, I-00146 Roma, Italy 
         \and
             Department of Physics and Astronomy, University of Padova, via Marzolo 8, I-35131 Padova, Italy
         \and
             Mullard Space Science Laboratory, University College London, Holmbury St. Mary, Surrey, RH5 6NT, United Kingdom
         \and
             INAF, Istituto di Astrofisica Spaziale e Fisica Cosmica Milano, via A. Corti 12, I-20133 Milano, Italy 
             }
   \date{Received August 8, 2021; accepted November 4, 2021}
\authorrunning{Krawczynski et al.}
  \abstract
{Phase-resolved  spectral and spectropolarimetric X-ray observations of magnetars present us with the opportunity to
 test models of the origin of the X-ray emission from these 
 objects, and to constrain the properties of the neutron
 star surface and atmosphere.}
 {Our first aim is to use archival {\it XMM-Newton} observations
 of the magnetar {\source} to ascertain how well four emission models
 describe the phase-resolved {\it XMM-Newton}
 energy spectra.
 Our second aim is to evaluate the scientific potential 
 of future spectropolarimetric observations of {\source} 
 with the {\it Imaging X-ray Polarimetry Explorer (IXPE)} 
 scheduled for launch in late 2021.
 The most salient questions are whether {\it IXPE} is able to
 distinguish between the different emission models, and whether
 {\it IXPE} can unambiguously detect the signatures
 of quantum electrodynamics (QED) effects in strong
 magnetic fields.}
{We used numerical radiation transport calculations for a large number of 
different system parameters to predict the X-ray flux and polarization energy 
spectra of the source {\source}. Based on the numerical 
results, we developed a new model to fit phase-resolved and phase-averaged X-ray 
spectral (i.e., {\it XMM-Newton} and {\it IXPE}) and spectropolarimetric ({\it 
IXPE}) data.
In order to test the sensitivity of {\it IXPE} to strong-field QED effects, we fit a simulated {\it IXPE} observation with two versions of the model, i.e., with and without QED effects accounted for.
}
{The fixed-ions condensed surface model gives the best description of the phase-resolved {\it XMM-Newton} spectra, followed by the blackbody and free-ions condensed surface models. 
The magnetized atmosphere model gives a poor description 
of the data and seems to be largely excluded. 
Simulations show that the {\it IXPE} observations of 
sources such as {\source} will allow us to cleanly 
distinguish between high-polarization (blackbody, 
magnetized atmosphere) 
and low-polarization (condensed surface) models. 
If the blackbody or magnetized atmosphere models apply, 
{\it IXPE} can easily prove QED effects based on 
$\sim$200\,ksec observations as studied here; 
longer {\it IXPE} observation times will be needed 
for a clear detection in the case of the condensed surface models.
}
{
The {\it XMM-Newton} data have such a good signal-to-noise ratio that they reveal some limitations of the theoretical models. Notwithstanding this caveat, the fits clearly favor 
the fixed-ions condensed surface and blackbody models
over the free-ions condensed surface and magnetized atmosphere models. 
The {\it IXPE} polarization information will greatly help us to figure
out how to improve the models. 
The first detection of strong-field QED effects in the signal 
from astrophysical sources seems possible if an adequate 
amount of time is dedicated to the observations.
}


   \maketitle
%

\section{Introduction}
   
   Magnetars are peculiar, isolated neutron stars (NSs)  believed 
   to host the strongest magnetic fields among astrophysical sources. Although historically they have been classified into two distinct classes, the anomalous X-ray pulsars (AXPs) and the 
   soft-gamma repeaters (SGRs), they share quite similar observational properties, including measured 
   spin periods $P$ in the $\sim 2$--$12$\,s range, 
   and period derivatives $\dot{P}$ in the 
   $10^{-13}$--$10^{-11}$\,s/s range. 
   Within the usual magneto-dipolar braking scenario, these values imply ultra-strong magnetic fields of up to 
   $B\approx10^{14}$--$10^{15}$ G, which is orders of magnitude greater than those derived in conventional rotation-powered pulsars. 
   
   Both AXPs and SGRs are observed to emit powerful bursts and flares in hard X-rays and in soft-$\gamma$ rays with peak luminosities 
   ranging from $10^{36}$ erg/s for the weaker and more frequent short bursts, and up to $10^{47}$ 
   erg/s for the rarest and most powerful giant flares \cite[detected only from three SGRs 
   up to now, see][]{maz+79,hur+99,pal+05}. The persistent emission in the $2$--$10$ keV band, $L_X\approx 10^{31}$-- $10^{36}$ erg/s (typically in excess of the rotational energy loss rate), is well fit by the superposition of a blackbody component ($kT\approx
   0.5$--$1$ keV) and a nonthermal, power-law tail, with photon index $\Gamma\approx2$--$4$, although  a purely thermal spectrum is observed in some (mostly transient) 
   sources. An additional power-law has been
   detected from several sources at higher energies \cite[$\gtrsim20$ keV, see][for reviews]{mer08,
   tzw15,kb17}.
   
   The most successful scenario in explaining the phenomenology of magnetars is the so-called
   twisted magnetosphere model \cite[][]{dt92,tlk02}. According to this model, the internal 
   magnetic field of the star is characterized by a nonnegligible toroidal component, which 
   is able to exert strong magnetic stresses on the conductive crust and displaces surface 
   elements to which the external magnetic field lines are anchored. As a consequence, the external dipole
   field  
   becomes ``twisted'', that is it acquires a sizable toroidal component. The presence of 
   such a toroidal component makes the external field nonpotential so that charged particles 
   must flow along the closed field lines.
   The situation thus differs from the case of conventional
   pulsars, where currents flow only along the open field lines \cite[][]{gj69}. 
   The presence of charges makes the (inner) magnetosphere optically-thick 
   to resonant Compton scattering (RCS). Comptonization of (thermal) radiation emitted by
   the NS surface results in the formation of the 
   power-law tail emission 
   observed in the soft X-ray band \cite[][]{tlk02,ntz08,zane+09}. Moreover, the plastic 
   deformations of the crust caused by the internal magnetic stresses may be also responsible 
   for the emission of short bursts and flares, through the injection in the magnetosphere 
   of an electron-positron fireball which eventually remains trapped within the closed field line 
   region \cite[][]{td95,td01,tt17}.

   For sources with ultra-strong magnetic fields, 
   such as magnetars, the polarization state of the 
   photons cannot be neglected. In these strong fields,
   photons propagate in two linearly polarized normal modes: 
   in the ordinary (O) mode, with the electric field oscillating in the plane 
   of the local magnetic field $\boldsymbol{B}$ and the propagation direction 
   $\boldsymbol{k}$; and in the extraordinary (X) mode, with the electric field oscillating 
   perpendicularly to both $\boldsymbol{k}$ and $\boldsymbol{B}$ \cite[see e.g.,][]{ginz70,
   gp74}. In such a strongly magnetized medium, the 
   properties of the radiative processes and opacities 
   depend strongly on the polarization state.
   The absorption of extraordinary photons for example
   is strongly suppressed outside the resonance 
   \cite[][]{pavl:76,her79,ven79}, 
   leading to a high polarization degree of the 
   RCS-reprocessed surface emission.
   The observed polarization fractions depend strongly on geometrical effects, which tend to depolarize the signal.
   For magnetic field strengths close to or exceeding 
   $B_{\rm q} \simeq 4.4\times 10^{13}$ G 
   QED effects play a role.
   The QED effects however tend to preserve the intrinsic polarization fraction \cite[see e.g.,][]{hsl03,tav+14,tav+15}.
   
   The spectral shape and polarization properties of the 
   persistent magnetar emission depend on the physical state of the neutron stellar surface. 
   The state of the surface is still a debated issue and no unanimous consensus has been reached as of yet.  In one model, a geometrically-thin/optically-thick atmosphere covers the surface of the star
   \cite[e.g.,][]{rom87,shib+92,pav+04,zavl+96,ls97,lai01}. 
   However, for field strengths and temperatures typical of magnetar sources,
   the surface layers may be in a condensed state \cite[][]{rud71,ls97,tur+04}, which would result in a 
   ``bare'', solid surface, so that surface emission is directly injected into the 
   magnetosphere. While X-ray spectral observations can test some predictions of these models, the spectral information alone is not sufficient to identify the emission model, as they tend to predict similar thermal energy spectra.
   The models however predict very different polarization properties \cite[see e.g.,][]{spw09,
   pot+12,tav+20}. As a consequence, X-ray polarimetry 
   can definitely probe the nature of magnetar surface layers.
   
   For these reasons, magnetars are among the primary targets 
   of the Imaging X-ray Polarimetry Explorer {\it IXPE} mission 
   scheduled for launch in 2021 \cite[][]{weiss+16}
   as well as the {\it eXTP} mission \cite[][]{zhang+19}. 
   The polarimetric observations will probe the emission 
   mechanism, and will provide new independent constraints 
   on the geometry of the source, removing degeneracies which 
   affect spectral analyses. Furthermore, the observations will
   allow us to observe the effect of 
   vacuum birefringence, a strong-field QED prediction \cite[][]{hsl03,tav+15,gonz+16,tav+20}.
   
   In this paper we focus on the AXP 1RXS J170849.0$-$400910 \cite[][]{isr+99}. The source is one 
   of the brightest (persistent) magnetar sources \cite[see the McGill online magnetar 
   catalog\footnote{http://www.physics.mcgill.ca/$\sim$pulsar/magnetar/main.html},][]{ok14} and is among the highest-priority
   targets to be observed in the first year of the {\it IXPE} science mission. 
   We first compare the energy spectra predicted by the magnetar emission models of \citet{tav+20} with archival {\it XMM-Newton} observations of this source \cite[see][]{rea+08,zane+09}.
   We perform for the first time not only a phase-averaged analysis, but also a phase-resolved analysis.
   For the best-fit models, we present simulated {\it IXPE} 
   observations, and show how well {\it IXPE} can distinguish between these emission models. 
   Our analysis indicates that the combined spectral and polarization data will allow us to identify the emission model, 
   and to constrain the source geometry and QED effects.
   After introducing the theoretical framework in Sect.\  \ref{s:theory}, we summarize the numerical implementation 
   in Sect.\  \ref{sec:numericalimpl}. 
   We discuss the implementation of the model as a 
   fitting model in Sect.\  \ref{s:sherpa}. 
   Sect.\  \ref{s:xmm} discusses the results from 
   fitting the {\it XMM-Newton} observations of {\source}. 
   Sect.\  \ref{s:ixpe} demonstrates what {\it IXPE} will 
   add to our studies of magnetars, by presenting the 
   analysis of simulated {\it IXPE} observations. 
   We close with a discussion of the results in Sect.\  \ref{s:disc}.
   
\begin{table*}[h!]
\caption{Parameters of the {\tt magnetar} model.\label{t:sherpa}}
\begin{center}
\begin{tabular}{lp{11.5cm}p{1.2cm}p{1.5cm}}
\hline
Parameter Name& Description& Units& Allowable Range\\ \hline \hline
{\tt model} & 1~(blackbody), 2~(magnetized atmosphere), 
3~(fixed-ions condensed surface), 4~(free-ions condensed surface) & - & 1-4 \\
{\tt norm} & overall normalization & - & 0 ... {\tt inf} \\
{\tt chi} & inclination from rotation axis & $[^{\circ}]$ & 0.1 ... 89.9 \\
{\tt xi} & 
angle between rotation and magnetic axes & $[^{\circ}]$ & 0.1 ... 89.9 \\
{\tt deltaPhi} & magnetic field line twist angle & [rad] & 0.3 ... 1.4   \\
{\tt beta} & electron bulk velocity & $[c]$ & 0.2 ... 0.7 \\
{\tt phi} & anticlockwise rotation to celestial north pole & $[^{\circ}]$ & 0 ... 180 \\  
{\tt phaseResolved} & 1 (0) for phase-resolved (phase-averaged) analysis  & - & 0, 1\\
{\tt offset} & offset added to the model phase (phase-resolved analysis only)& - & 0. ... 1. \\ 
{\tt dir} & +1 (-1) for pulse evolution as modeled (inverted) (phase-resolved analysis only) & - & -1, 1 \\ 
{\tt phase1} & start of interval for averaging (phase-averaged analysis only)&-& 0 ... 1\\
{\tt phase2} & end of phase interval for averaging, a value smaller than {\tt phase1} will result in averaging the model from {\tt phase2} to {\tt phase1} (phase-averaged analysis only) &-& 0 ... 1\\ \hline
\end{tabular}
\end{center}
\end{table*}   
\section{Theoretical framework}\label{s:theory}
   \subsection{Magnetospheric structure} \label{subsec:magnetosphere}
    In this and the following sections, we recall the most important underlying assumptions we used
   in our work to produce the model archive for fitting the observational and
   simulated data.
      In the framework of the twisted magnetosphere model \cite[][]{tlk02}, the amount of shear
      of the external magnetic field is usually 
      described through the so-called twist angle $\Delta\phi$, which measures the 
      angular displacement between the footpoints of a given field line. Although 
      the twist is more likely restricted to localized bundles of field lines 
      \cite[][]{tlk02,bel09}, we assumed for the sake of simplicity that the external 
      dipolar field is globally twisted: namely, all the external field lines are 
      sheared by the same amount $\Dfns$, maintaining the global axial symmetry 
      \cite[][see also \citealt{ntz08,fd11}]{tav+14}.We stress that the magnetic field topology surrounding an ultra-magnetized NS is likely to be much more complicated than that of a
      (twisted) dipole. Small-scale magnetic flux tubes may rise close to the star surface, as suggested, for example, by the detection of a phase-dependent (proton) cyclotron line in the X-ray spectrum of SGR 0418-5729 \cite[][]{tiengo13}. However, multipolar components decay 
      much faster with increasing distance from the star than the dipole component. Given that resonant scatterings mostly occur at about $10\,\Rns$ \citep{tlk02,ntz08}, only the dipole component of the field survives in the region of interest. 
      
      Under these conditions, 
      the polar components of the star magnetic field can be expressed as
      \begin{equation} \label{eqn:Btwisted}
          \boldsymbol{B}=\frac{\Bp}{2}\left(\frac{r}{\Rns}\right)^{-p-2}\left[-\frac{\der{f}}
          {\der\cos\theta},\frac{pf}{\sin\theta},
          \sqrt{\frac{Cp}{p+1}}\frac{f^{1+1/p}}{\sin\theta}\right]\,,
      \end{equation}
      where $\Bp$ is the polar magnetic field strength, $\Rns$ the star radius and $f$ 
      is a function of the magnetic colatitude $\theta$, obtained as a solution of the 
      Grad-Shafranov equation \cite[see][]{tlk02,pav+09}. The radial index $p$ and the 
      parameter $C$ are related to the twist angle $\Dfns$ by
      \begin{equation} \label{eqn:Dfns}
          \Dfns=\sqrt{\frac{C}{p(p+1)}}\lim_{\theta\rightarrow0}\int_{\theta}^{\pi/2}
          \frac{f^{1/p}}{\sin\theta}\der\theta\,.
      \end{equation}
      Since $C$ is an eigenvalue of the problem, it is completely determined once $p$ is assigned. This implies that the globally twisted field is fixed once $\Bp$ and $p$, or equivalently $\Dfns$, are provided.
      The latter option is adopted in this paper.

      It has been shown that the magnetospheric currents, which must stream along
      the closed field lines due to the additional toroidal component in the external 
      field, should be sustained mainly by electron-positron 
      pairs
      \cite[][]{bt07}. However, since a detailed model which accounts for pairs has not yet been developed, we consider the simplified scenario in which 
      the charge carriers are electrons and ions extracted from the crust due to 
      the strong surface magnetic field \cite[][]{ft07,ntz08,fd11,tav+14}. Ions, however, 
      are lifted at much smaller heights above the surface (as they are much heavier 
      than electrons) and they are expected to contribute less to the emerging photon spectrum than electrons \cite[see][for a detailed 
      analysis]{tlk02}. Following these considerations, we adopt the 
      uni-directional flow approximation, in which only electrons are considered
      to stream along the closed field lines. In our simulations, we described the 
      motion of these magnetospheric electrons as a bulk motion (from the northern to 
      the southern magnetic hemisphere) at constant velocity $\beta$ in units of the speed 
      of light. Moreover, in order to account for the velocity spread, we superimposed a (relativistic) Maxwellian distribution at the temperature
      $T_{\rm e}$, which is one-dimensional because of the magnetic
      confinement which affects electrons
      perpendicularly to the $B$-field \cite[see also][for more details]{ntz08}, to the bulk motion. Clearly, the assumption of constant bulk velocity is an oversimplification. Charges accelerate after they are extracted from the star surface, reach a maximum velocity, and then decelerate as they hit the outermost layers. This was discussed by \citet{bt07} in connection with a simplified, but more realistic, model in which pair creation is accounted for. Again, much as in the case of the dipole field assumption, we are mostly concerned with what happens at about ten star radii, far from the initial or terminal parts of the charge trajectory. In this respect taking the electron velocity to be a constant is not entirely unrealistic.
      Under these assumptions, the density of magnetospheric particles can be
      expressed in terms of the magnetic field strength and the twist parameter 
      as \cite[][]{tlk02,ntz08}
      \begin{equation} \label{eqn:edens}
          n_{\rm e}=\frac{p+1}{4\pi e}\frac{B}{\beta r}\left(\frac{B_\phi}{B_\theta}\right)\,,
      \end{equation}
      where $e$ is the electron charge. This value turns out to be sufficiently high to make the medium optically thick for RCS. As the photon energy $E$ reaches in the particle rest 
      frame the electron cyclotron energy $\omega_{\rm ce}=\hbar eB/\me c\simeq 
      11.6\,(B/10^{12}\,{\rm G})$ keV (with $\me$ the electron mass), the photon 
      is absorbed by the electron, which is in turn excited to its first Landau 
      level. The particle de-excitation, however, occurs in a very short time 
      \cite[$\approx 3\times10^{-14}\, (B/10^{11}\,{\rm G})^{-2}\,{\rm s}$, see 
      e.g.][]{fd11} and a photon is emitted at the same energy, so that the process 
      is akin to a scattering event in all respects. In the case of magnetars, 
      with surface radiation emitted at $E\approx 1$ keV and $B_{\rm p}\sim10^{14}$ 
      G, the resonance condition is met as the magnetic field strength has dropped 
      down to $\sim10^{11}$ G, which occurs at about $5$--$10\,\Rns$. Due to RCS, 
      thermal photons emitted from the stellar surface are expected to be up-scattered 
      by magnetospheric electrons, populating the nonthermal tail in the soft 
      X-ray spectra ($0.1$--$10$ keV) of these sources \cite[see e.g.][for reviews]{mer08,
      re11,tzw15,kb17}.
      
\begin{table*}[h!]
\caption{Results from fitting the {\it XMM-Newton} EPIC-pn data with the four phase and angle averaged absorbed emission models.\label{t:xmmA}}
\begin{center}
\begin{tabular}{|p{4cm}|c|c|c|c|c|}
\hline
Model Name& 
$\chi^2$/DoF & 
{\tt norm} &
 $\Delta\phi$ & 
 $\beta$ &
 $n_{\rm H}$  \\
 & & [$10^{-8}$] & [rad] & [$c$] & [$10^{22}{\rm cm}^{-2}$]\\
 \hline
Blackbody             & 1122.8/67 & 12.3 & 0.30 & 0.30 & 0.494 \\ 
Magnetized atmosphere      & 39239/67 &  4.9 & 0.30 & 0.20 & 0.537\\ 
Free-ions cond.\,surf. & 2372.6/67 & 10.2 & 0.30 & 0.34 & 0.471  \\ 
Fixed-ions cond.\,surf.& 858.85/67 & 9.4 & 0.30 & 0.33 & 0.295 \\  \hline
\end{tabular}
\end{center}
\end{table*}
\begin{figure*}[h!]
\centering
\includegraphics[width=0.45\linewidth]{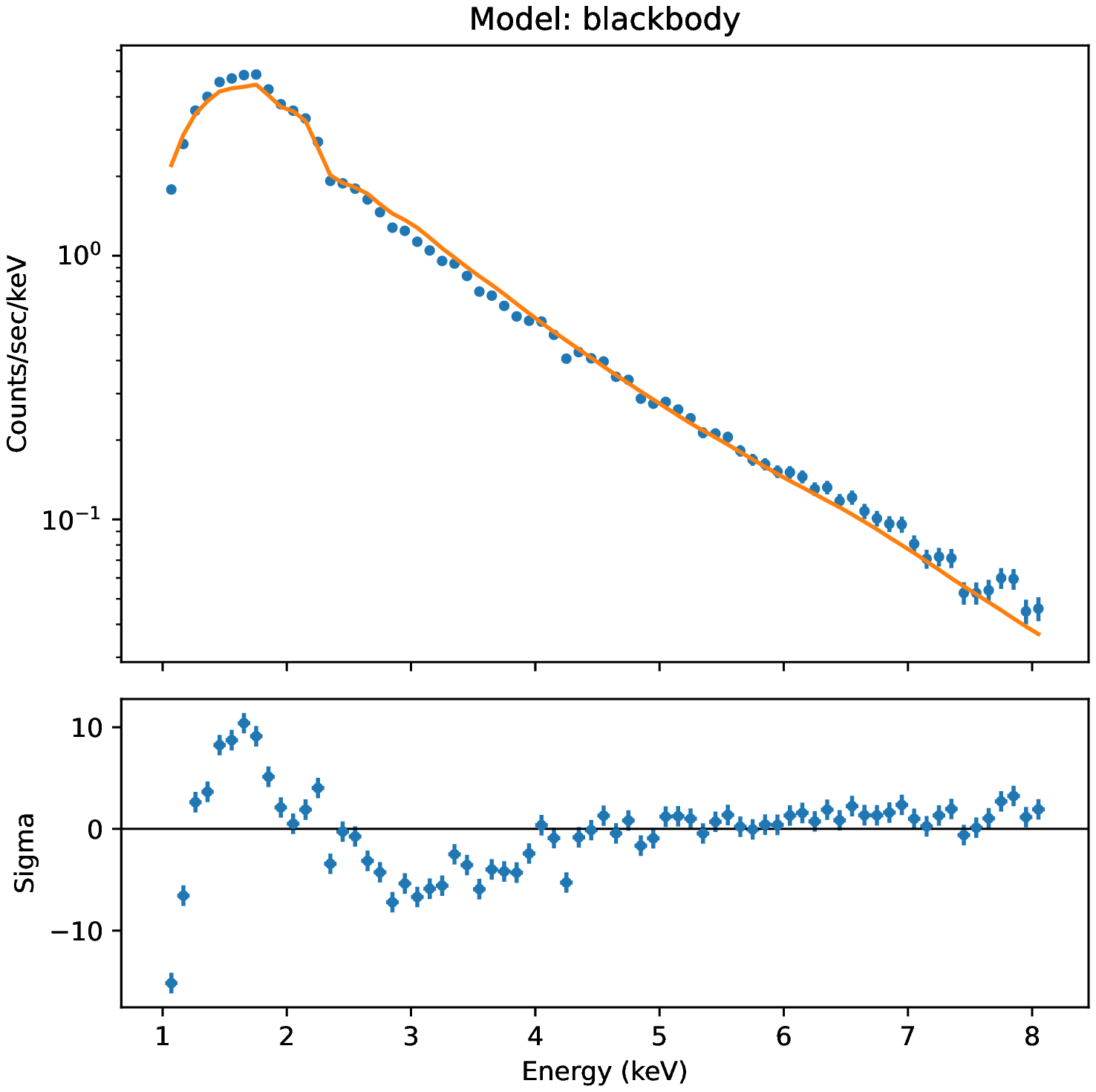} \hspace*{0.5cm}
\includegraphics[width=0.45\linewidth]{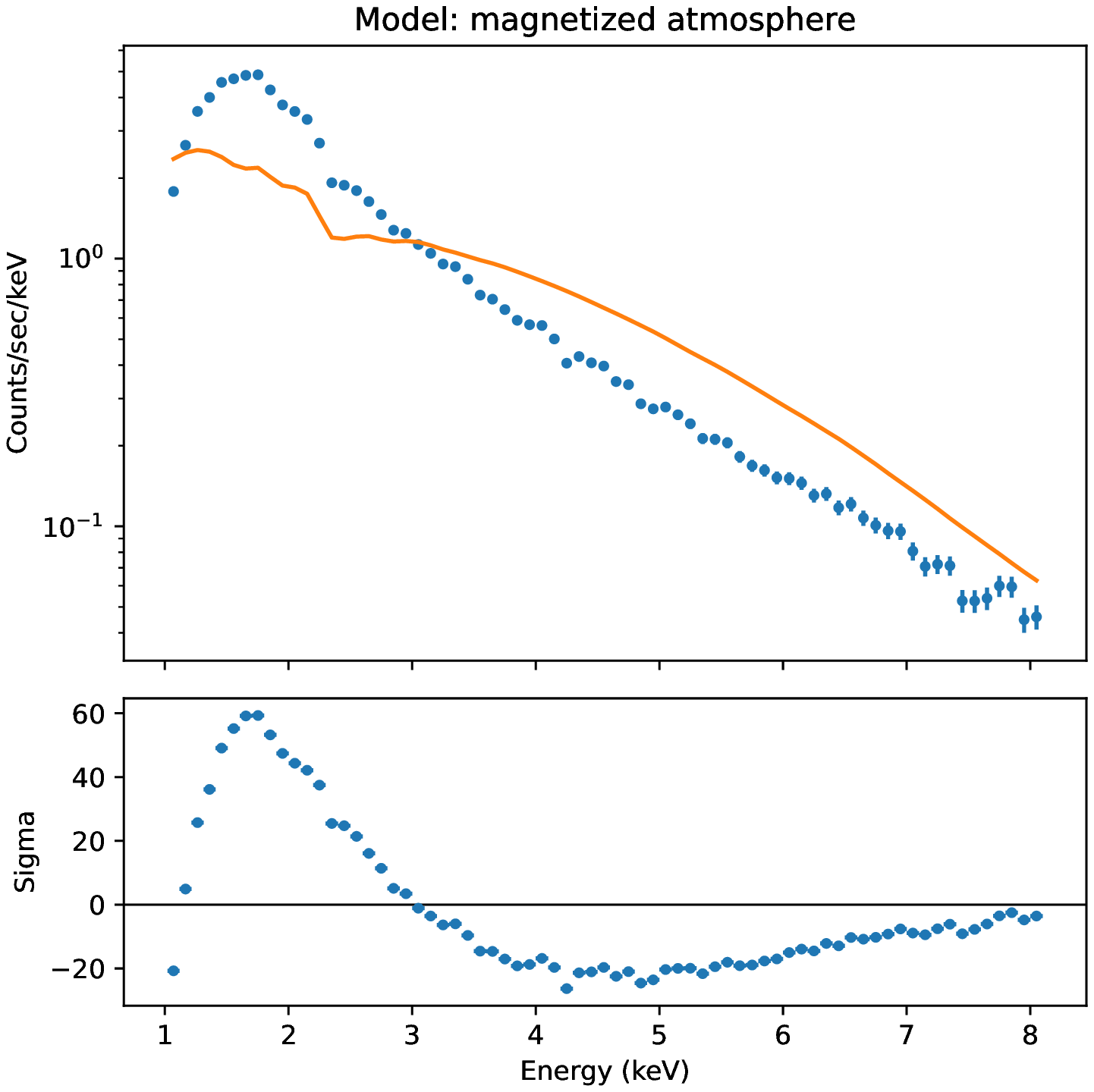}
\includegraphics[width=0.45\linewidth]{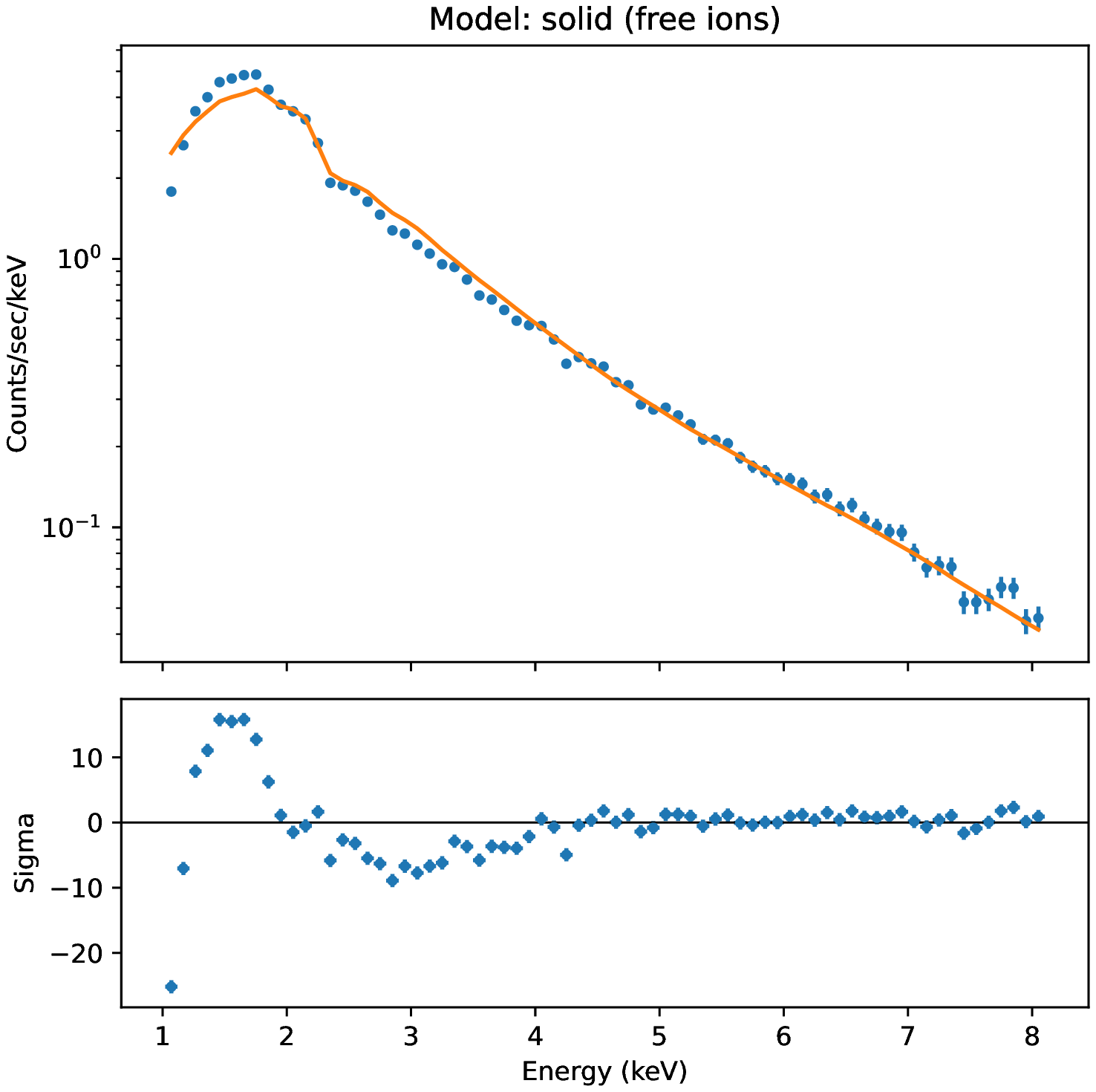}\hspace*{0.5cm}
\includegraphics[width=0.45\linewidth]{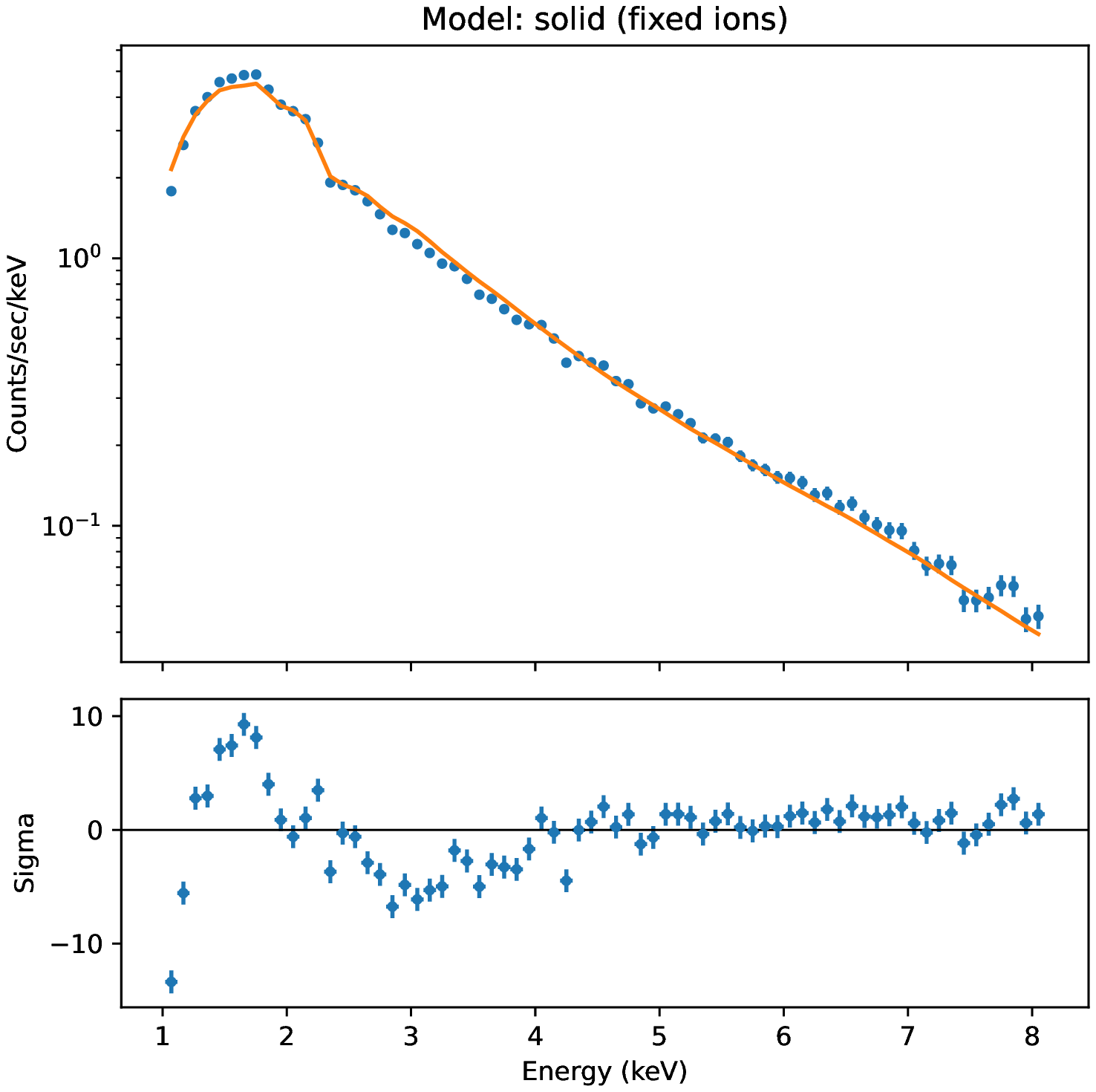}
\caption{Best-fit models of the phase-averaged {\it XMM-Newton} data
(upper panels, blue data points) with the phase 
and angle averaged models (upper panels, orange lines).
The lower panels show (data-model)/sigma for the spectra shown above. 
\label{f:XMMspecA}} 
\end{figure*}
   \subsection{Photon polarization transport} \label{subsec:polarization}
      In the strong field limit the cross sections which describe the 
      interactions between photons and charged particles can be dramatically different 
      with respect to the unmagnetized ones, and they change according 
      to the photon polarization state. Strong magnetic fields 
      have a strong impact on the polarization dependent 
      cross sections of radiative processes such as Compton scattering 
      \cite[or even second-order processes such as free-free emission and photon splitting, 
      see][]{lieu81,lauer+83,ston79,bul98}, and the scatterings are more strongly suppressed for X-mode photons than for O-mode ones 
      \cite[][]{lai+10}. RCS is not an exception; photons originally polarized
      in the O-mode that resonantly scatter off electrons may change their polarization 
      state into the X-mode and vice versa as given by the RCS cross sections 
      \cite[][]{tlk02,ntz08,tav+14}:
      \begin{equation} \label{eqn:RCScrosssection}
         \begin{array}{lll}
            \sigma_{\rm O-X}&=&3\sigma_{\rm O-O} \\
            \ &\ &\ \\
            \sigma_{\rm X-O}&=&\dfrac{1}{3}\sigma_{\rm X-X}\,.
         \end{array}
      \end{equation}
      Equations (\ref{eqn:RCScrosssection}) imply that 
      photons emerging from the magnetosphere are preferentially polarized in the 
      extraordinary mode, with a typical ratio of 3:1 with respect to
      those emerging in the ordinary mode. 
      
\begin{table*}[h!]
\caption{Results from fitting the {\it XMM-Newton} EPIC-pn data with the four phase-resolved absorbed magnetar emission models.\label{t:xmm}}
\begin{center}
\begin{tabular}{|p{4cm}|c|c|c|c|c|c|c|c|}
\hline
Model Name& 
$\chi^2$/DoF & 
{\tt norm} &
$\chi$  & $\xi$  & 
 $\Delta\phi$  & 
 $\beta$&
 $n_{\rm H}$ &   {\tt offset}  \\
 
 & & [$10^{-9}$] & [$^{\circ}$] & [$^{\circ}$] & [rad] & [$c$] &  [$10^{22}{\rm cm}^{-2}$]  & 

\\ \hline
Blackbody             & 3372.5/703 & 12.6       & 0.1 &0.1& 0.40 & 0.47 & 0.610 & 0.864\\ 
Blackbody$^{\dagger}$         & 3419.4/703 & 12.1         & 15 & 60 & 0.30 & 0.40 & 0.556 & 0.860\\ 
Magnetized atmosphere      & 37432.5/703 &  6.14  & 15.0 & 0.1 & 0.30 & 0.20 & 0. & 0.856\\ 
Free-ions cond.\,surf. & 3728.9/703 & 9.16 & 0.1 & 87 & 0.30 & 0.39 & 0.335& 0.864  \\ 
Fixed-ions cond.\,surf.& 3103.9/703 & 8.74  &15 & 60 &0.50 & 0.35 & 0.513 & 0.858\\  \hline
\end{tabular}
\\ $^{\dagger}$ Fit with $\chi$ and $\xi$ tied to the best-fit values of the fixed-ions condensed surface model.
\end{center}
\end{table*}
\begin{figure*}[h!]
\centering
\includegraphics[width=0.45\linewidth]{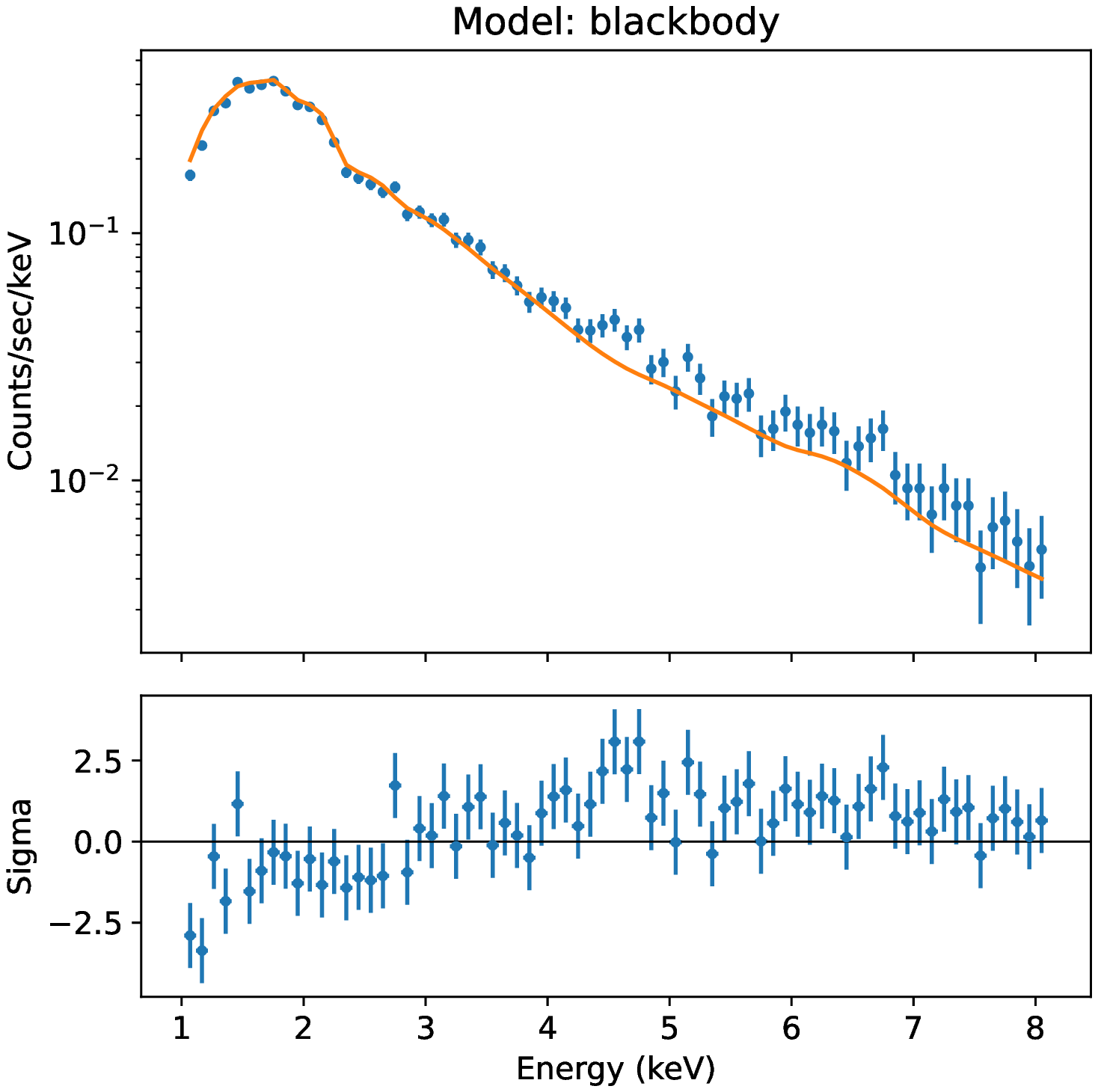} \hspace*{0.5cm}
\includegraphics[width=0.45\linewidth]{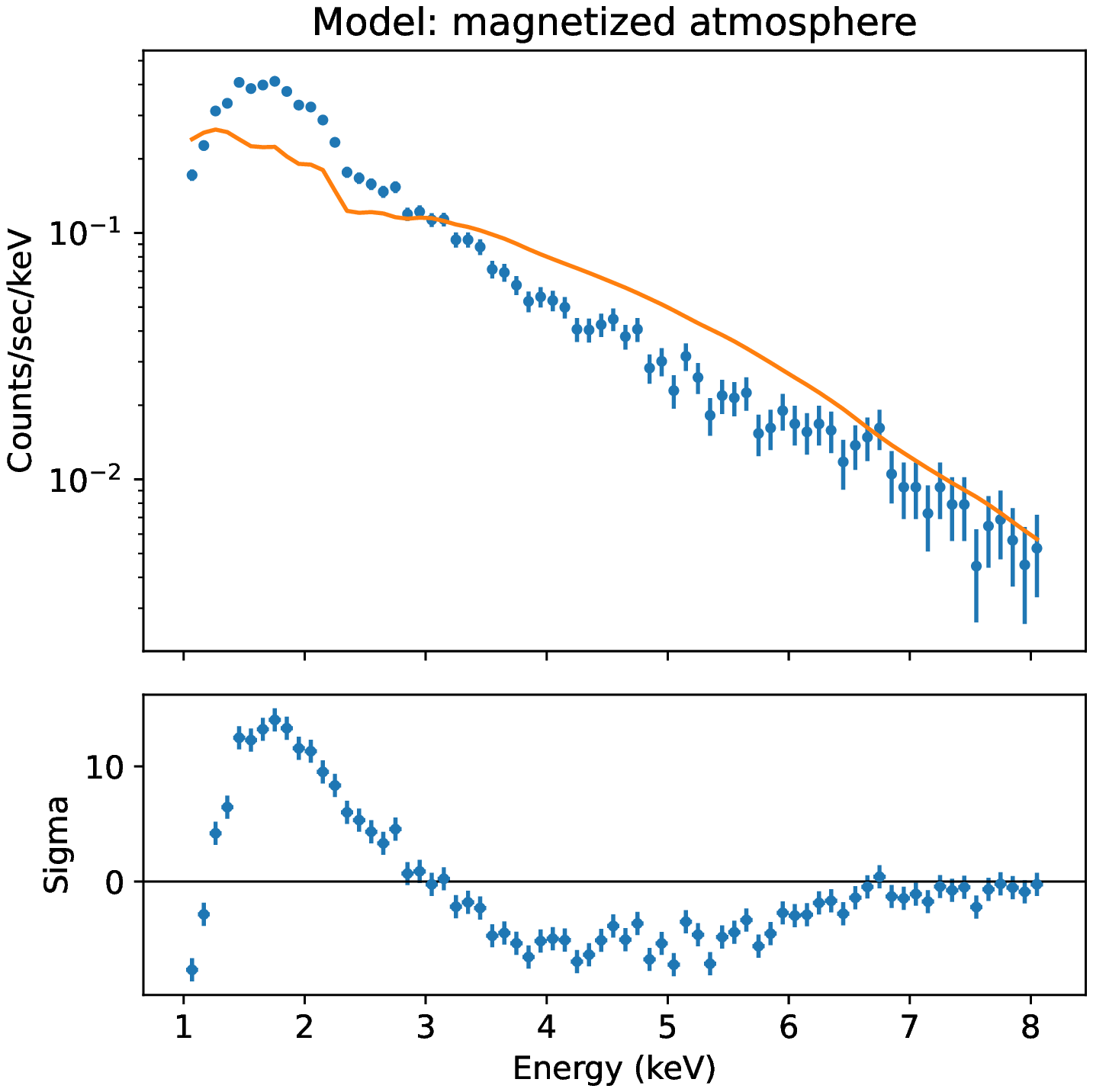}
\includegraphics[width=0.45\linewidth]{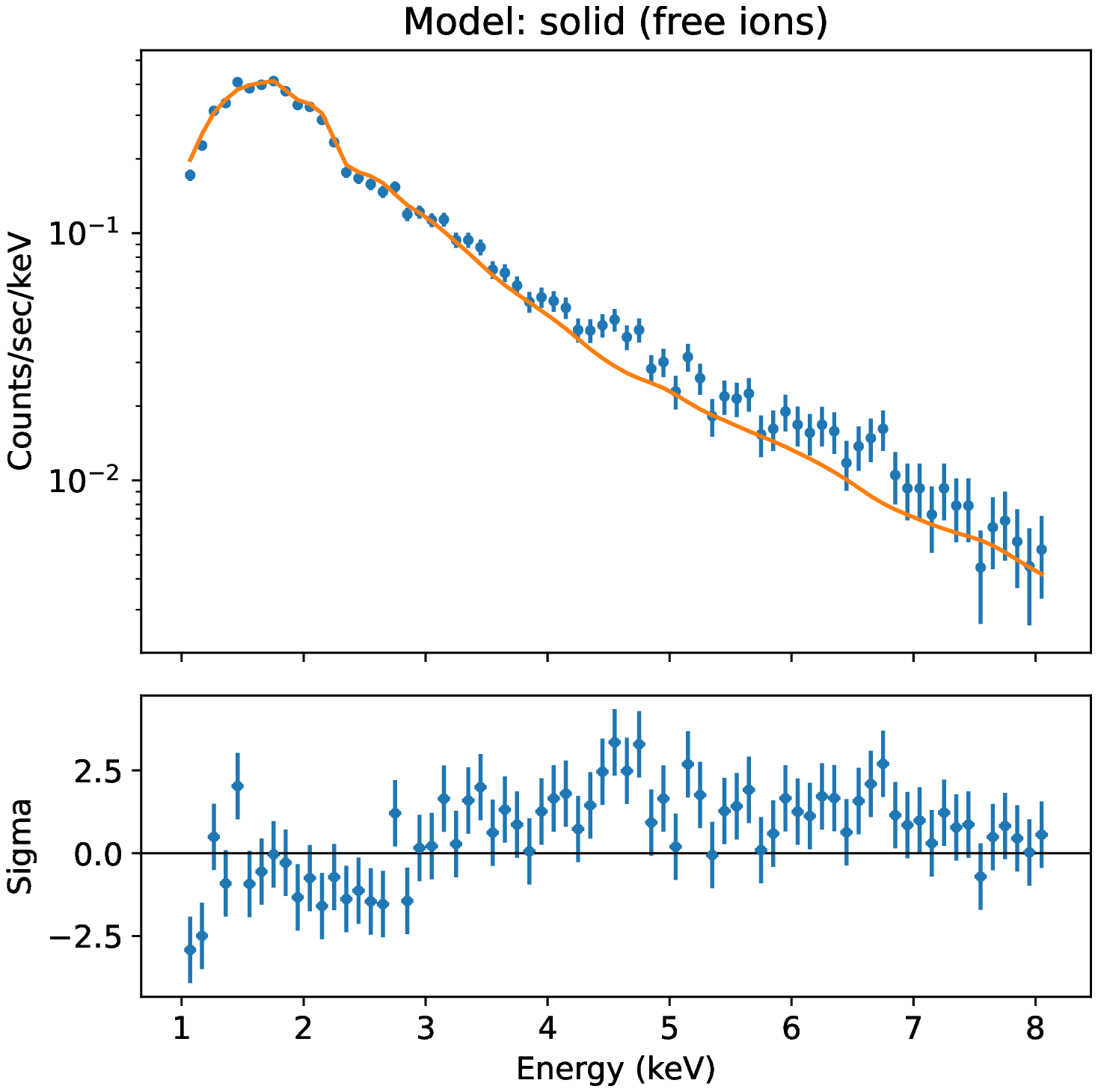}\hspace*{0.5cm}
\includegraphics[width=0.45\linewidth]{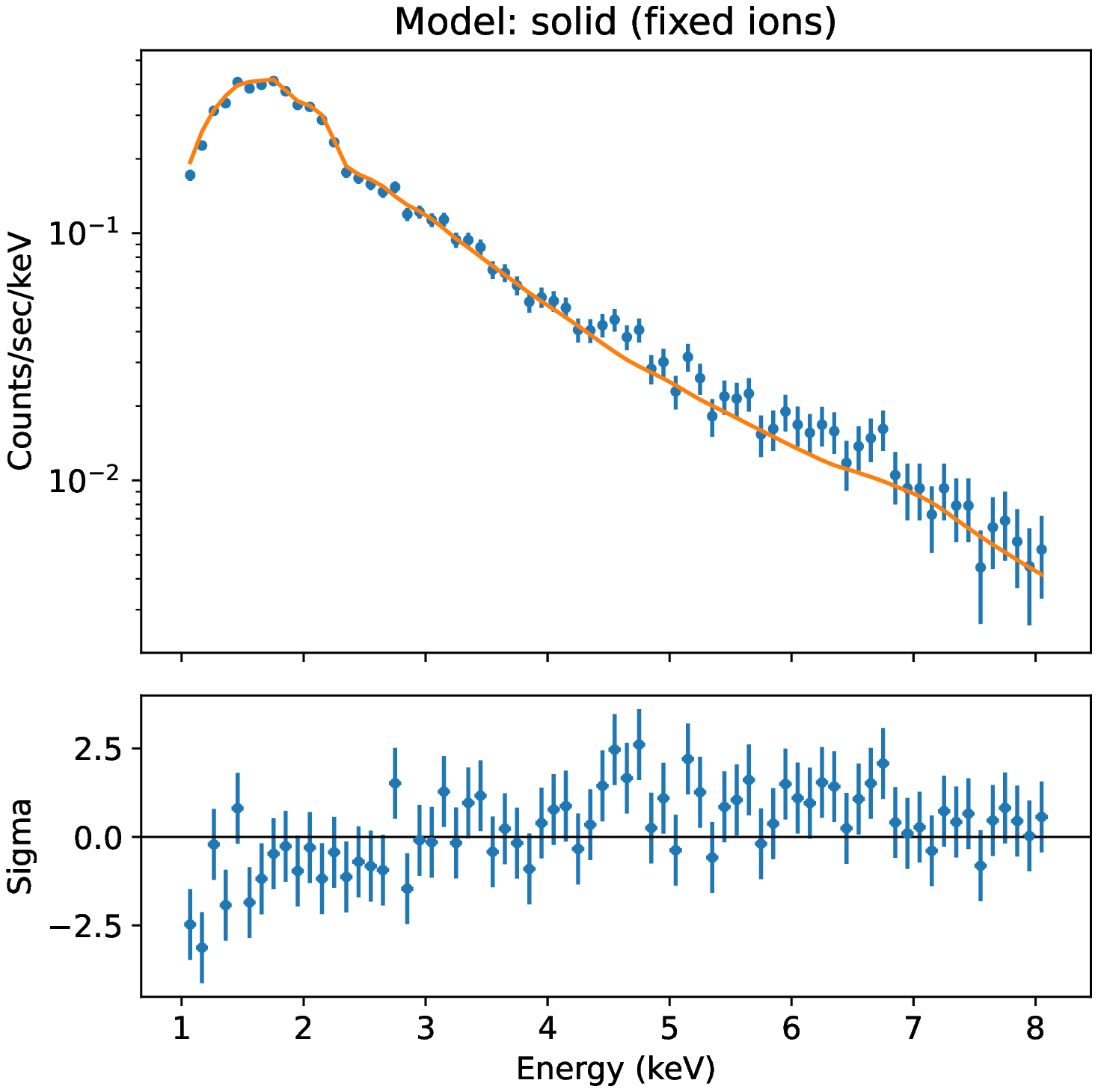}
\caption{
Exemplary spectra of the {\it XMM-Newton} data for phase bin 3
(upper panels, blue data points) and models (upper panels, orange line), that is
blackbody model (top left), 
magnetized atmosphere model (top right), 
free-ions condensed surface (bottom left), and
fixed-ions condensed surface (bottom right).
The lower panels show (data-model)/sigma for the spectra above.
\label{f:XMMspec}}
\end{figure*}
\begin{figure*}[h!]
\centering
\includegraphics[width=0.45\linewidth]{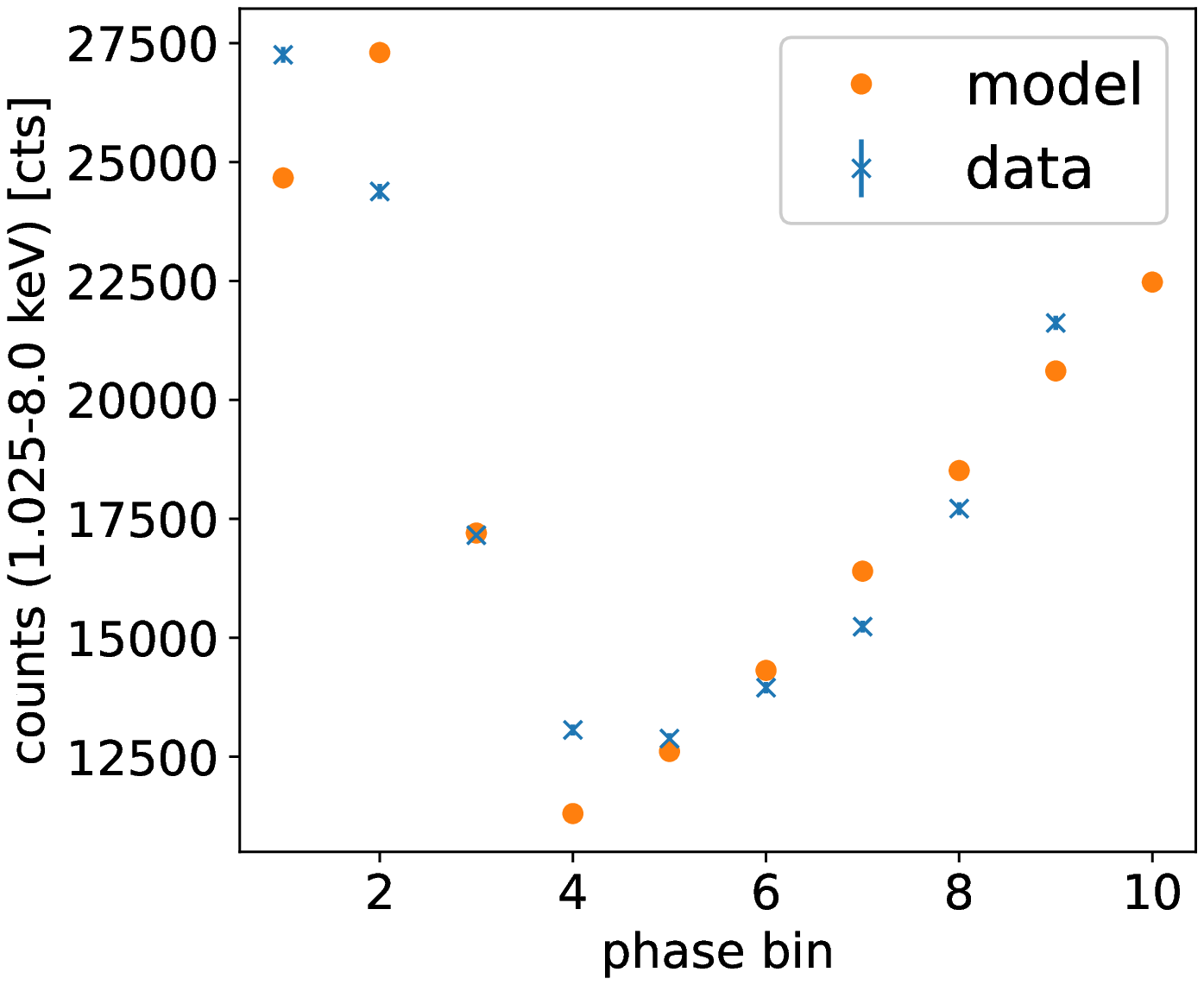}\hspace*{0.5cm}
\includegraphics[width=0.45\linewidth]{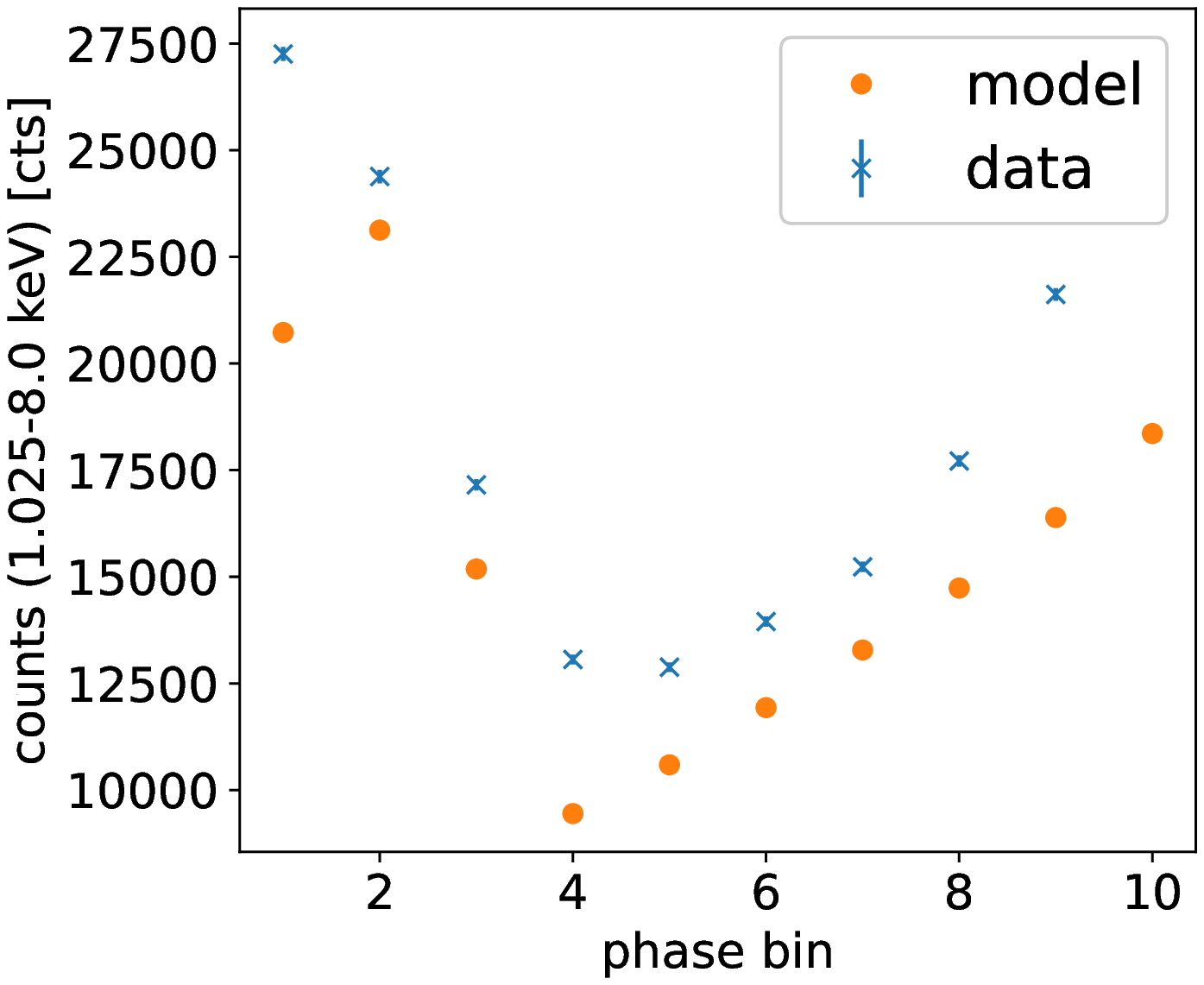}
\includegraphics[width=0.45\linewidth]{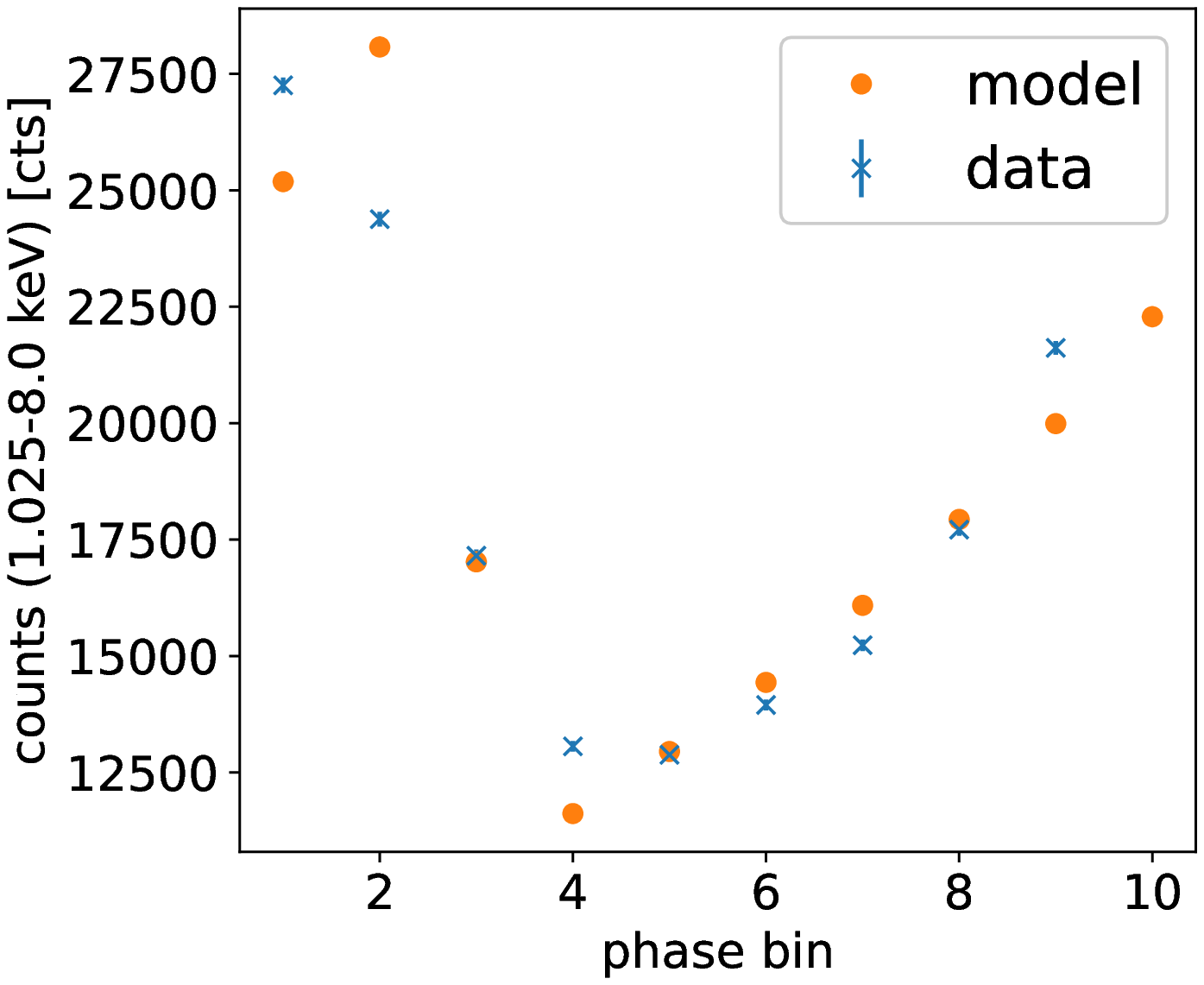}\hspace*{0.5cm}
\includegraphics[width=0.45\linewidth]{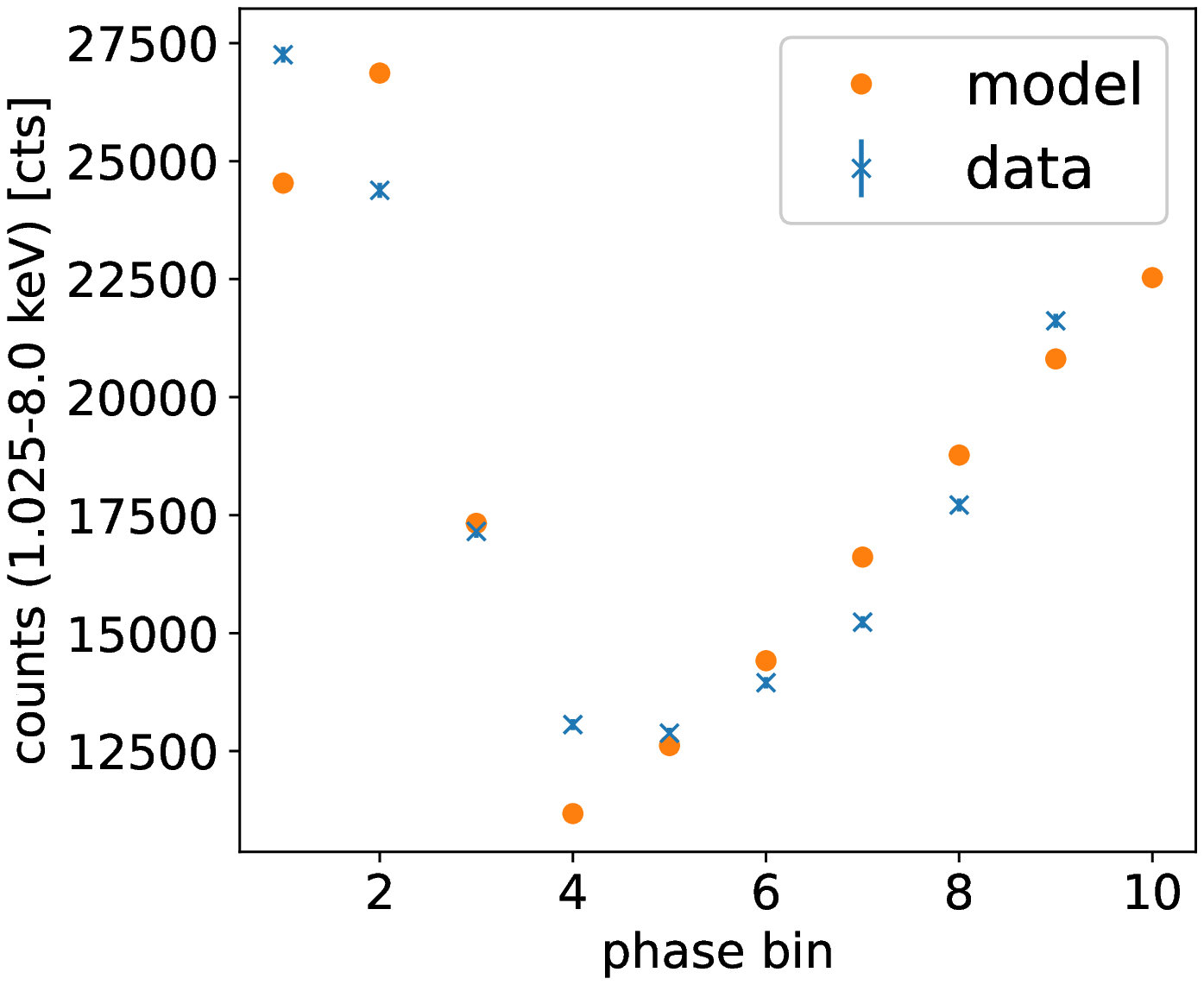}
\caption{Observed (blue crosses) and theoretical (orange circles) 
0.5-8\,keV pulse diagrams of
the best-fit blackbody model (top left),  
magnetized atmosphere model (top right), 
free-ions condensed surface (bottom left), and
fixed-ions condensed surface (bottom right).
The {\it XMM-Newton} data are shown with error bars which are so small 
that they cannot be seen for most data points.
\label{f:lc1}}
\end{figure*}
\begin{figure*}[h!]
\centering
\includegraphics[width=0.95\linewidth]{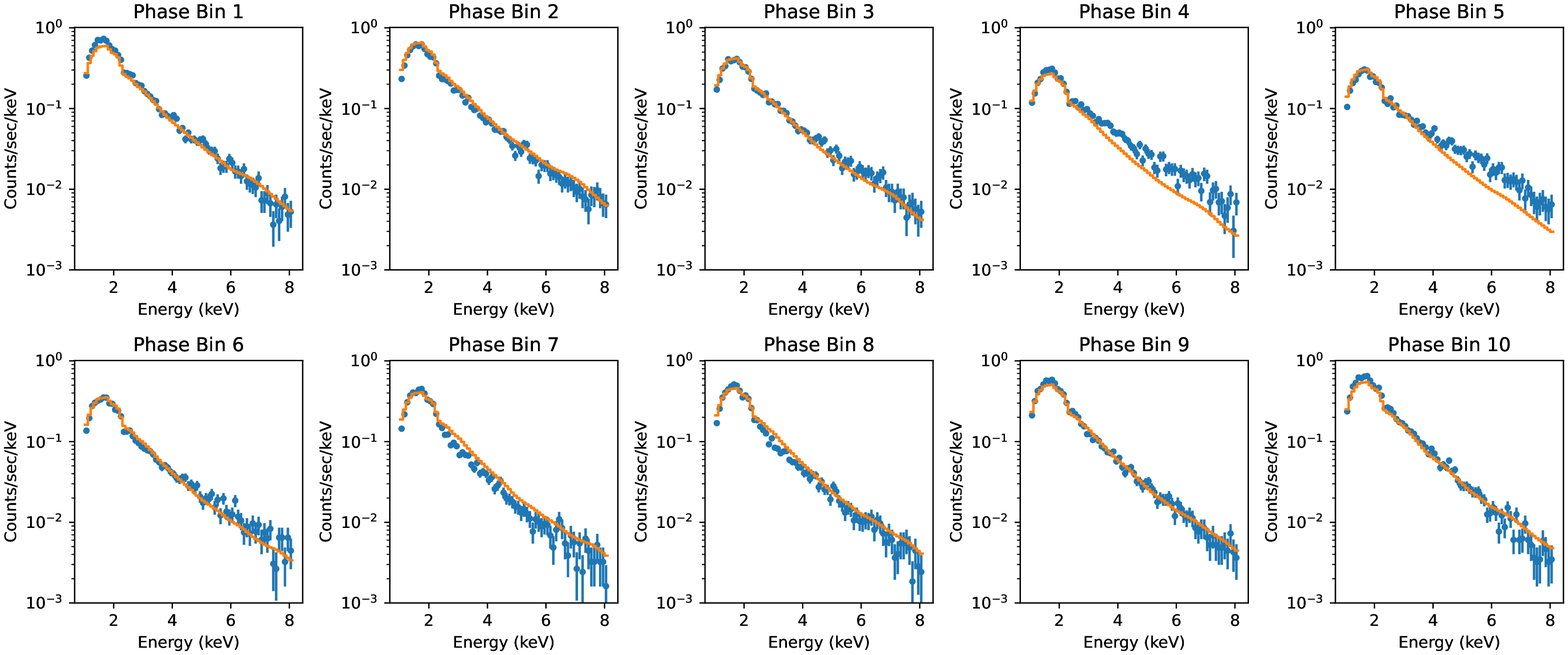}
\includegraphics[width=0.95\linewidth]{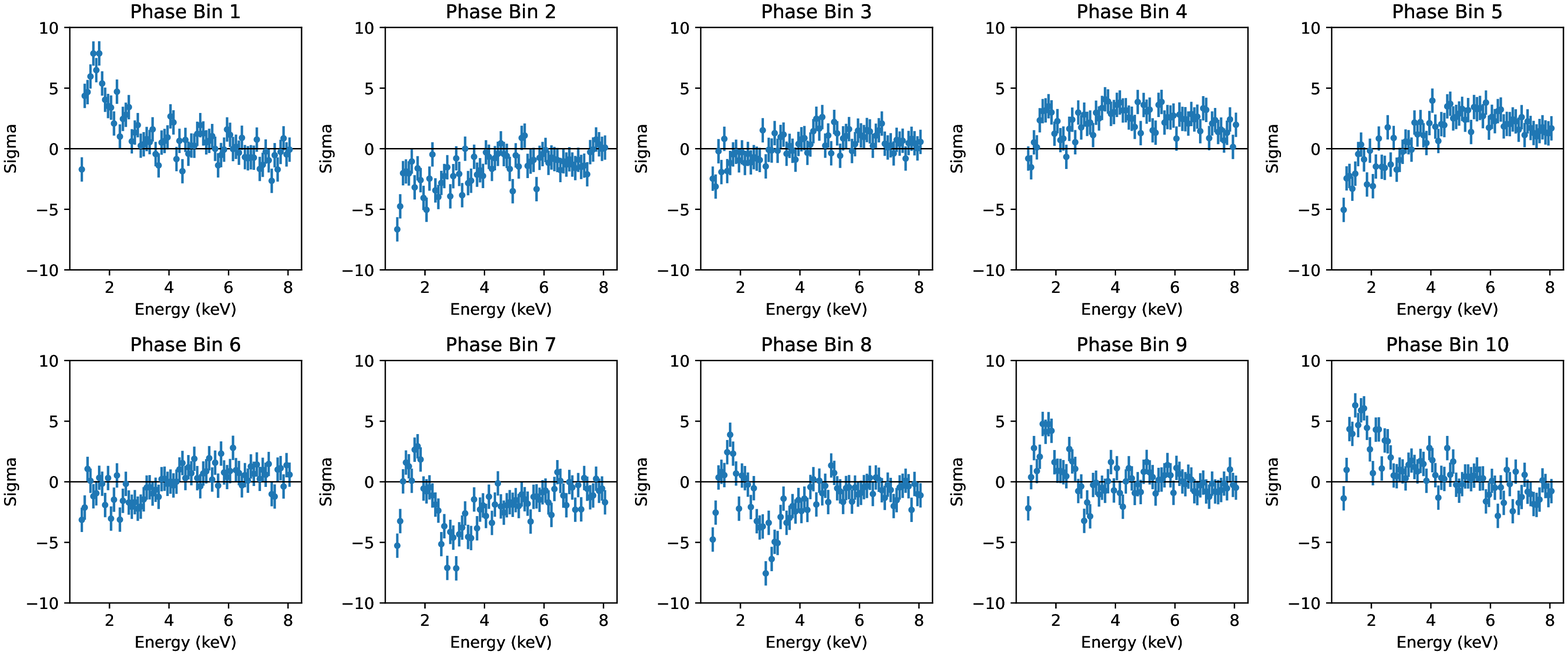}
\caption{Upper panels: {\it XMM-Newton} data 
of all ten phase-resolved spectra (blue data points) 
and the best-fit model
(fixed-ions condensed surface, orange lines). 
Lower panels: deviation ((data-model)/sigma) for all
spectra.
\label{f:all3}}
\end{figure*}
      In the case of magnetars, however, photon polarization states can be also 
      modified as they propagate in vacuo due to vacuum polarization, a 
      QED effect originally theorized by \citet{he36} and 
      independently by \citet{weiss36}. Ultra-strong magnetic fields (typically 
      in excess of the quantum critical field $\Bq\simeq 4.4\times10^{13}$ G) can 
      polarize the virtual electron-positron pairs which populate the magnetized 
      vacuum around the star. As a consequence, the components of both the 
      dielectric and magnetic permeability tensors of vacuum deviate from unity, 
      ensuring that O- and X-mode photons propagate with slightly different 
      refraction indices (vacuum birefringence). In order to understand how QED effects
      can influence the polarization properties of the emitted radiation, one should
      solve the wave equation, accounting for the vacuum polarization terms.
      It turns out that, close to the surface, the strong magnetic field of the star
      can force the photon polarization vector to maintain its original 
      orientation relative to the magnetic field direction (adiabatic propagation), which in turn can vary along the 
      photon trajectory. Hence, the polarization pattern of the emitted radiation is
      practically unchanged until the star magnetic field becomes too weak to affect 
      the photon polarization states. This typically occurs at a distance $r_{\rm pl}$
      from the star, called the polarization-limiting radius \cite[see][]{hsl03,tav+15}, 
      \begin{equation} \label{eqn:adiabaticradius}
      r_{\rm pl}\simeq 4.8\left(\frac{B_{\rm p}}{10^{11}\,{\rm G}}\right)^{2/5}
      \left(\frac{E}{1\,{\rm keV}}\right)^{1/5}\left(\frac{\Rns}{10\,{\rm km}}\right)^{1/5}\,\Rns\,.
      \end{equation}
      This radius increases with the magneic field strength.
      As the photons move outward, the polarization vectors freeze. The photon polarization states can deviate from the initial ones if the magnetic 
      field direction still changes substantially along the photon trajectory. As a consequence of these effects, stronger magnetic fields lead to polarization signatures that are closer to those at emission, allowing us to probe the physical processes at the surface and in the inner magnetosphere 
      through polarimetric measurements \cite[see e.g.,][]{tav+15,gonz+16}.
      
      Finally, another important effect may take place when plasma contributions to the 
      dielectric tensor become comparable to those of vacuum polarization \cite[the 
      so-called vacuum resonance, see][]{lh02,lh03}. This occurs at a density
      \begin{equation}
          \rho_{\rm V}\simeq 0.964Y_{\rm e}^{-1}\left(\frac{E}{1\,{\rm keV}}\right)^2 
          \left(\frac{B}{10^{14}\,{\rm G}}\right)^2\zeta^{-2}\,{\rm g\,cm^{-3}}\,,
      \end{equation}
      where $Y_{\rm e}$ is the plasma electron fraction and $\zeta$ is a slowing varying
      function of the magnetic field strength. In these conditions photon polarization
      modes can switch from O to X and vice versa, with a probability
      \begin{equation}
          P_{\rm con}=1-\exp\left[-\frac{\pi}{2}\left(\frac{E}{E_{\rm ad}}\right)^3\right]\,,
      \end{equation}
      where $E_{\rm ad}$ is the energy at which mode conversion occurs.
      
   \subsection{Surface emission models} \label{subsec:emissionmodels}
      As described in \citet{tav+20},  
      we consider three different models to predict 
      the spectral and polarization 
      properties. In the first model, photons are emitted 
      directly from the stellar surface following a blackbody distribution and they
      are assumed to be 100\% polarized in the extraordinary mode. In the second
      model, photons are reprocessed in an optically-thick, geometrically-thin magnetized 
      atmosphere that covers the stellar surface. Finally, in the third model, emission 
      occurs from the ``bare'', solid surface of the star, which is assumed to be
      composed by a magnetic condensate. For ease of reading, we briefly summarize 
      in the following the main properties of each emission model, referring the 
      reader to \citet{tav+20} for more details.
      \subsubsection{Blackbody model}
         The currently available soft X-ray spectra of many AXPs and SGRs can be fit well with the superposition of a thermal (blackbody) component 
         and a nonthermal power law tail \cite[see][for
         reviews]{mer08,tzw15}. 
        The power-law component can be identified with photons, resonantly up-scattered by magnetospheric 
         electrons \cite[see e.g.,][compare Sect.\  \ref{subsec:magnetosphere}]{tlk02,ft07,ntz08}. The thermal component can be associated with the cooling emission from the stellar surface. Following these observational 
         findings, we make the zero-order approximation 
         to describe photons from the stellar surface as
         blackbody emission at the temperature inferred from 
         the observed spectra (that is $T\approx0.5$ keV for our template source). As illustrated in section \ref{subsec:polarization}, 
         the ultra-strong surface magnetic field of the star is expected to strongly 
         affect the polarization pattern of the emitted radiation, favoring the propagation 
         of extraordinary photons with respect to ordinary ones. On the other hand, 
         some models predict that thermal O-mode photons are more likely emitted from 
         hot-spots on the surface where magnetospheric currents return, which are more 
         concentrated in the (magnetic) equatorial regions \cite[see][]{tlk02,bt07,fd11}. 
         A self-consistent model that includes 
         the effects of returning currents has not yet been
         developed. We thus make the simplified assumption that
         the blackbody emission is 100\% polarized in the extraordinary mode.
         
      \subsubsection{Magnetized atmosphere}
         The presence of a magnetized atmospheric layer (with typical scale-height 
         $0.1$--$10$ cm) above the surface of ordinary pulsars and even highly-magnetized 
         isolated NSs has been suggested by many authors \cite[see e.g.,][]{ls97,lai01,zp02}
         and different models have been developed to describe the
         radiation emitted from such atmospheres, both in the magnetized ($B\ga10^{12}$
         G) and in the nonmagnetized limits \cite[see][for a complete list of 
         references]{spw09}. However, model atmospheres around passively cooling NSs, which are in thermal and hydrostatic equilibrium, might not apply to active magnetars because the structure of such an atmosphere
         would be strongly affected by the bombardment of returning charged particles flowing along the closed field lines.
         Despite recent attempts to investigate the properties of
         such bombarded atmospheric layers \cite[see][]{gonz+19}, a complete understanding of the emitted spectrum and polarization is still lacking. 
         With the aim to provide a more physically motivated emission model alongside the 100\% polarized blackbody model mentioned above, we consider a simple plane-parallel
         magnetized atmosphere and use the radiation 
         transfer code of \citet{spw09} to characterize the 
         energy spectrum and polarization properties of the
         emitted radiation. The code solves the radiative transfer equation for the ordinary and the extraordinary photons in the case of a partially ionized, 
         pure hydrogen atmosphere using 
         the opacities described in \citet{pch14}. 
         The polarization properties of the emitted radiation are then computed 
         self-consistently from the O-mode and X-mode specific intensities  \cite[see][for more details]{tav+20}. 
         In this respect, we remark that the mode conversion induced by the vacuum 
         resonance (see Sect.\  \ref{subsec:polarization}) is taken into account as discussed in \citealt[see also \citealt{vl06}]{lh03}.

      \subsubsection{Condensed surface}
         For values of the magnetic field strength in excess of $\sim2.4\times10^9$
         G such that the electron gyro-radius is comparable to the Bohr radius,
         the properties of matter are strongly modified. In particular, due to the 
         confinement in the direction perpendicular to the magnetic field, the atomic 
         electron clouds are squeezed in the direction of $\boldsymbol{B}$ and, for 
         temperatures below a value $T_{\rm crit}$ \cite[that depends on the 
         composition, see][]{lai01,ml07}, molecular chains can be formed via covalent 
         bonding. Under these conditions, a phase transition occurs \cite[the so-called 
         magnetic condensation,][]{rud71,ls97}, and the stellar surface turns into a 
         ``naked'' solid surface \cite[][]{tur+04}. Considering the values of the 
         surface temperature and the spin-down magnetic field strength inferred 
         from observations \cite[][]{isr+99}, 
         it turns out that magnetic condensation is indeed a plausible scenario for 
         {\source} \cite[see for example Fig.\  1 in][]{tav+20}. In order to reproduce the 
         spectral and polarization properties of the emitted radiation in this case, 
         we resort to the approximate fitting formulae developed by \citet[see also 
         \citealt{gonz+16,tav+20}, for more details]{pot+12}, that compute the 
         emissivities for ordinary and extraordinary photons for a selected chemical 
         composition (we adopted a Fe condensed surface). In our simulations we provide 
         results in the two limits of (i) free-ions, that is ions 
         on the stellar surface are free to move in response to the electromagnetic waves, 
         and (ii) fixed-ions, with ions considered as fixed in a lattice. As stated 
         in previous works \cite[see e.g.,][]{vad+05}, the real situation should probably 
         lie in between these two limits. We finally remark that, much in the same
         way as in the blackbody and in the magnetized atmosphere cases, we neglect the effects 
         of returning currents in the condensed surface model as well.
      
\section{Numerical implementation} \label{sec:numericalimpl}
   As described in \citet{tav+20}, we produced a number of different simulations using the Monte Carlo
   code originally developed by \citet{ntz08}  with the addition of a specific module to account for
   photon polarization properties \cite[see][]{tav+14}. The code starts by dividing the 
   stellar surface into a number of equal-area patches, each labeled by the magnetic 
   colatitude $\theta$ and azimuth $\phi$ of its center. Seed photons are launched randomly from each of these patches, according to the selected emission 
   model. This can easily be achieved for the blackbody model using a series expansion method \cite[see e.g.,][]{bc70}. 
   In the case of the magnetized atmosphere and condensed surface models, we use an acceptance/rejection method \cite[][]{vonn51,press+92}.
   A similar sampling method is used to determine the 
   polarization state of the emitted photons\footnote{The only exception is the blackbody
   model, for which, as mentioned in Sect.\  \ref{subsec:emissionmodels}, photons are
   considered to be 100\% polarized in the X-mode.}. 
   The number $N_{\theta,\phi}$ of seed photons 
   which are launched from each surface patch is determined by setting a reference number 
   $\bar{N}$ for the patch which emits the smallest number of photons and weighing the photons of all the other patches according to the considered emission model. In our 
   calculations, we chose a $\theta\times\phi=10\times10$ angular mesh for the blackbody 
   and the condensed surface models, and a $20\times10$ angular mesh for the magnetized atmosphere model, 
   choosing  $\bar{N}$ in such a way that the total number of photons 
   launched from the surface is $\sim 10^7$ for all the emission models. Photons are eventually collected on a $15\times15$ angular mesh on the sky at infinity.  
   
   The code follows each photon along its trajectory and accounts for the resonant scatterings off magnetospheric electrons.
   In the adiabatic region close to the stellar surface,
   vacuum polarization effects are properly 
   considered by locking the polarization mode to that set at  emission or to the polarization mode resulting after each
   scattering. 
   The wave equation is then solved for the photon Stokes parameters $I$, $Q$, $U$ and $V$ 
   in the region between the RCS last scattering radius $r_{\rm esc}$ and the polarization
   limiting radius $r_{\rm pl}$ \cite[see][for more details]{tav+14,tav+20}. We neglect strong gravity effects such as relativistic ray-bending to save computational resources, which allows us to run the simulations for each emission 
   model in $\approx 5$ days on a single-CPU computer. 
   We remark, however, that the effects of relativistic ray-bending 
   should be small, as scatterings occur 
   at a large distances from the star (see Sect.\  \ref{subsec:magnetosphere}), where general 
   relativistic effects are weak.
   Strong gravity effects do not play a major role 
   for the polarization state of photons propagating 
   around magnetars. In fact, the typical length scales 
   along which the photon polarization vectors rotate due to strong-field QED effects (see Sect.\  
   \ref{subsec:polarization}) are much smaller 
   than those relevant for general 
   relativistic effects \cite[see][]{cs77,sc77,cps80}.
 
\begin{figure*}[h!]
\centering
\includegraphics[width=0.45\linewidth]{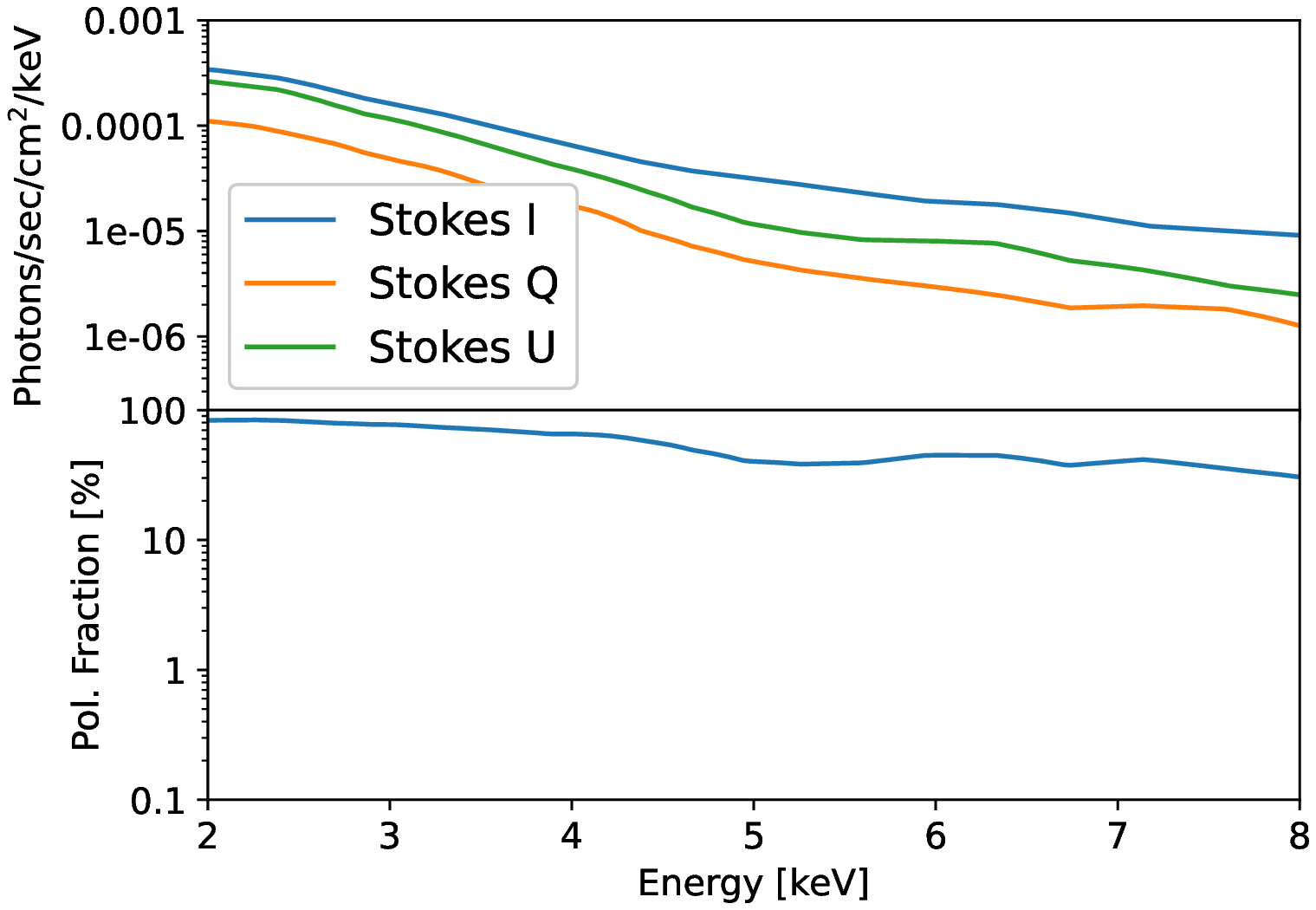}\hspace*{0.5cm}
\includegraphics[width=0.45\linewidth]{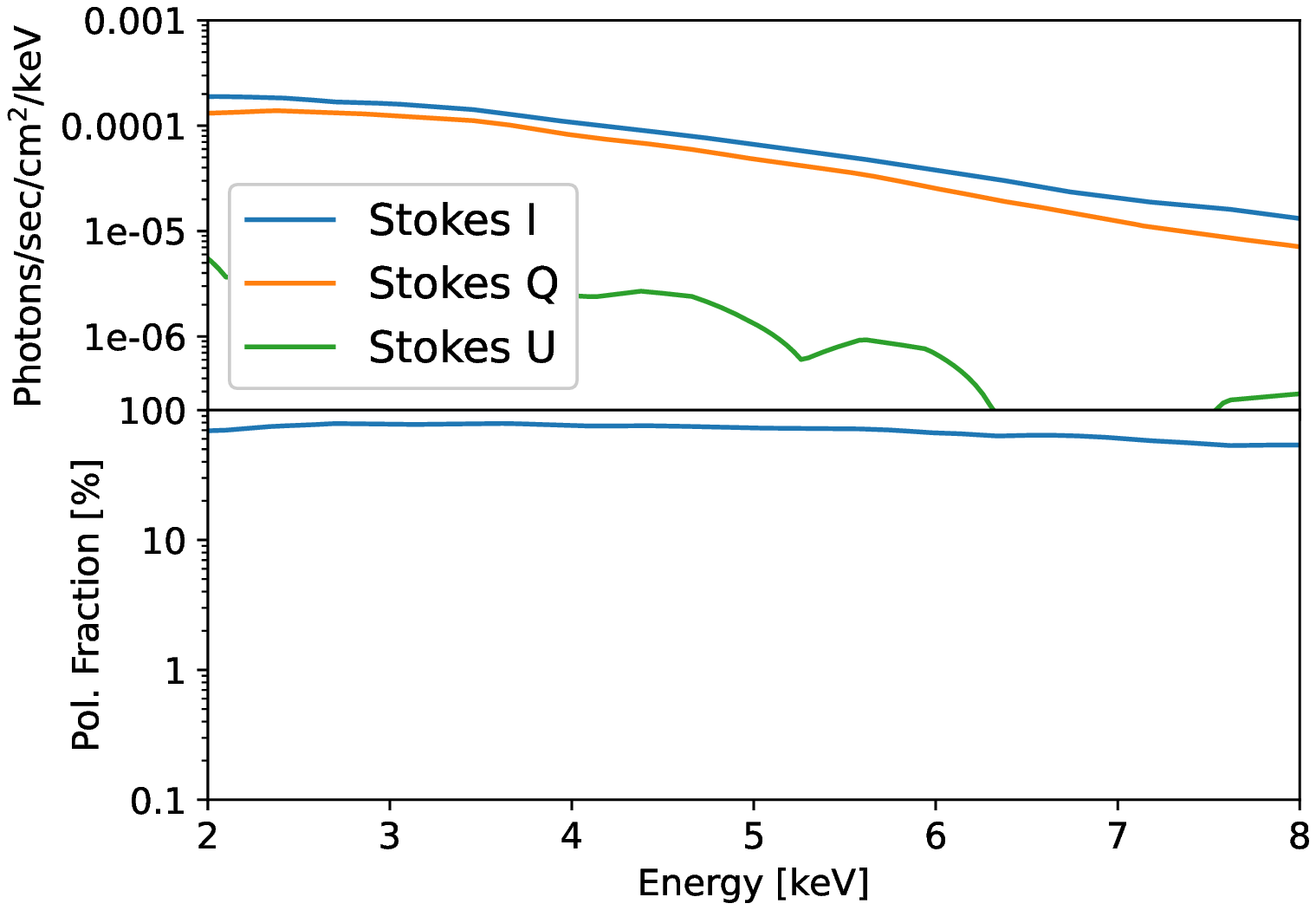}
\includegraphics[width=0.45\linewidth]{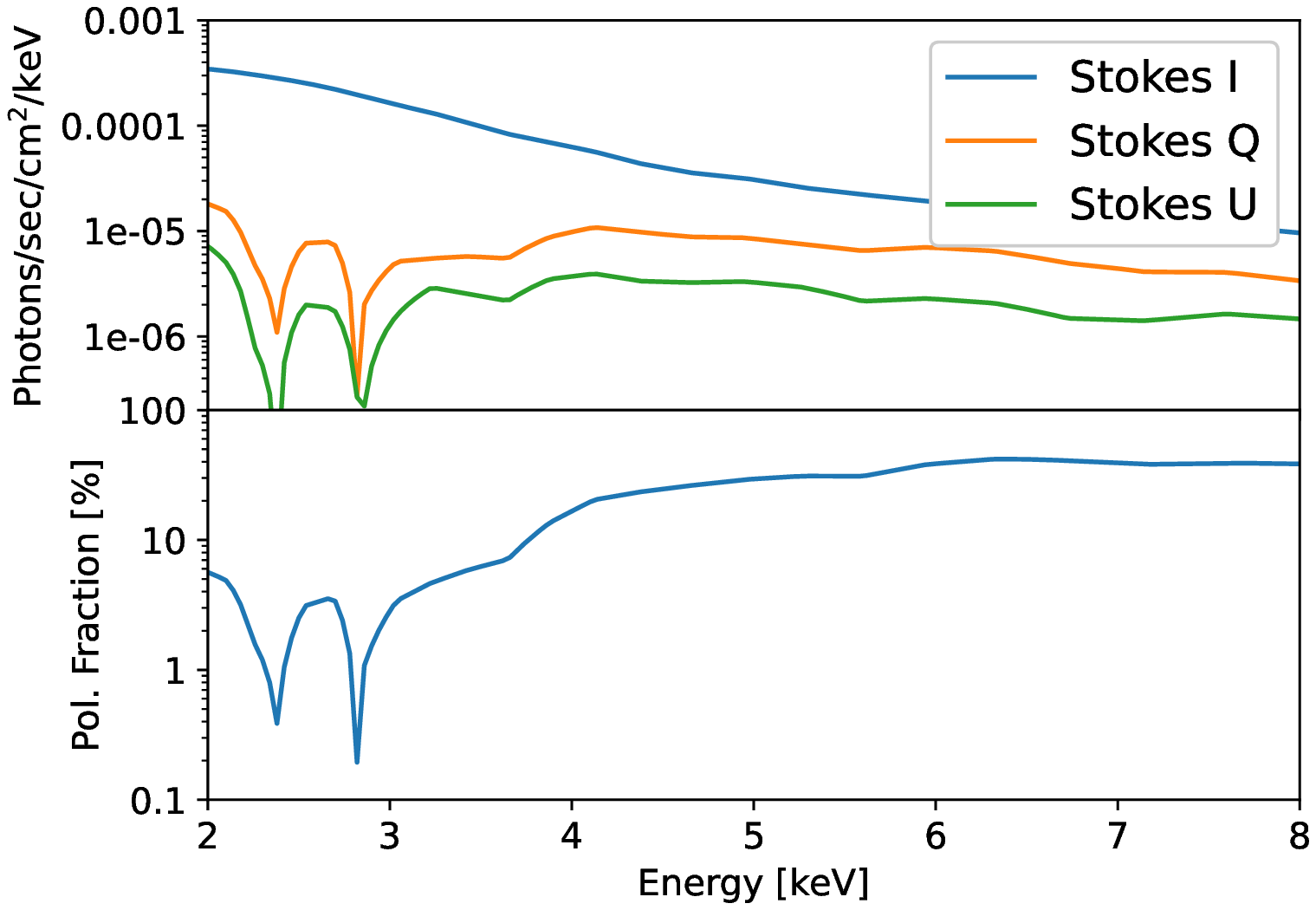}\hspace*{0.5cm}
\includegraphics[width=0.45\linewidth]{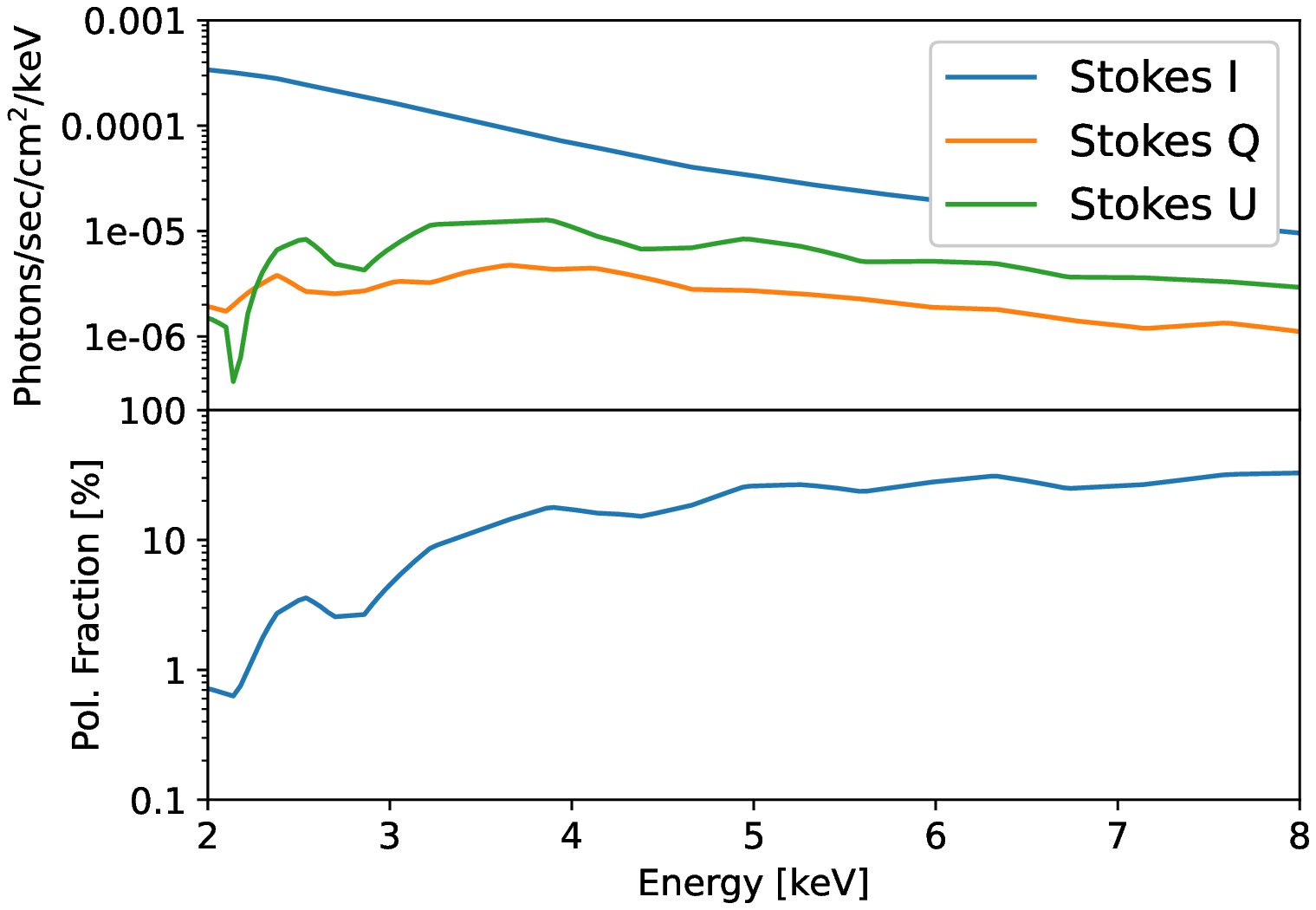}
\caption{Absolute value of Stokes I, Q and U parameters (top panels) and
linear polarization fraction (lower panels) for 
the blackbody model (top left), magnetized atmosphere model (top right), 
free-ions condensed surface (bottom left), and
fixed-ions condensed surface (bottom right).
For all models, the results for phase bin 3 are shown.
\label{f:polMod}}
\end{figure*}

   We run the Monte Carlo code assuming a fixed value of the polar magnetic field strength
   $\Bp=5\times10^{14}$ G, which is compatible with {\source} and with many of the known
   AXPs and SGRs. For all the emission models, we assume a constant 0.5 keV surface temperature \cite[see][for a complete discussion
   on this choice]{tav+20}. We simulate twist angles
   $\Dfns$ and charge bulk velocities $\beta$ between $0.3$--$1.4$ rad (step size: $0.1$ rad) and $0.2$--$0.7$ (step size: $0.1$), respectively. 
   We added the value $\beta=0.34$ to the grid, corresponding to the electron bulk velocity obtained from the spectral
   fitting by \citet{zane+09}. 
   We assume the same electron temperature $T_{\rm el}=10$ keV 
   for all cases. Each single run is post-processed 
   with an {\sc idl} script to derive the energy spectra seen
   by an observer at infinity for different values of the two angles $\chi$ and $\xi$, which measure the inclination of the line-of sight and of the magnetic axis wrt the star rotation axis, respectively.
   The outputs of this script consist in one ascii file for each value of $\chi$, $\xi$, $\Dfns$ and $\beta$. $\chi$ and $\xi$ 
   range between $0$--$180^\circ$ (step $15^\circ$) and $0$--$90^\circ$ (step $15^\circ$), 
   respectively. Each output file contains the Stokes parameters $I$, $Q/I$ and $U/I$ 
   as functions of the photon energy (sampled by a 49-bin grid between $0.5$ and $10$ keV) 
   and the rotational phase (sampled by a 9-bin grid between $0$ and $2\pi$).
      
\section{Loading the model in {\tt sherpa}}\label{s:sherpa}
The spectral analysis is based on fitting the observed and simulated 
$I$ ({\it XMM-Newton} and {\it IXPE}) and $Q$ and $U$ ({\it IXPE}) energy 
spectra with a least squares technique based on the forward folding method
\cite[][]{2015APh....68...45K,2017ApJ...838...72S,2020KrawczKerrC}.
We used the {\tt kerrC} model developed for fitting the 
spectropolarimetric observations of black holes 
as a template to develop a new model, called {\tt magnetar}, 
to work with the data tables described above.
The model uses the spectral fitting package {\tt sherpa}, 
the general purpose fitting model developed as part 
of the {\tt CIAO} {\it Chandra} analysis software
\citep{2001SPIE.4477...76F,2007ASPC..376..543D}.
 The fitting module is written in {\tt Python 3.7} 
and uses the {\tt astropy} library  for saving and storing 
the predicted Stokes parameter $I$, $Q$, and $U$ 
energy spectra in the {\tt fits} format, 
the {\tt numpy} library for fast manipulation 
of the spectral data, and the {\tt scipy} 
library for the interpolation between simulated parameter values
using the {\tt regularGridinterpolation} tool.

\begin{table*}[h!]
\caption{Results from fitting the simulated {\it IXPE} blackbody data. \label{t:bb}}
\begin{center}
\begin{tabular}{|p{4.5cm}|c|c|c|c|c|c|c|c|c|}
\hline
Model Name& 
$\chi^2$/DoF & 
{\tt norm} &
$\chi$  & $\xi$  & 
 $\Delta\phi$  & 
 $\beta$&
 $n_{\rm H}$ &  $\phi$ 
 &  {\tt offset}  \\
 & & [$10^{-9}$] & [$^{\circ}$] & [$^{\circ}$] & [rad] & [$c$] &  [$10^{22}{\rm cm}^{-2}$] & [$^{\circ}$] & 
\\ \hline
Blackbody (input)   & NA       & 12.1   & 15.0   &60.0 & 0.30 & 0.40 & 0.556 & 0 &0.860\\ 
Blackbody (best fit)& 195.0/202& 12.1         & 15.7 &59.8 & 0.33 & 0.37 & 0.557& -1.7 & 0.860 \\ 
Fixed-ions cond.\,surf. (best fit)& 754.6/202   &9.26&15.7 & 60.0 &0.30 & 0.44 & 0.642& 1.1 & 0.919\\  
Blackbody (no QED)& 711.3/202& 12.2         & 15.7 &59.8 & 0.33 & 0.37 & 0.557& -1.7 & 0.860 \\ \hline
\end{tabular}
\end{center}
\end{table*}
\begin{figure}[h!]
\centering
\includegraphics[width=0.95\linewidth]{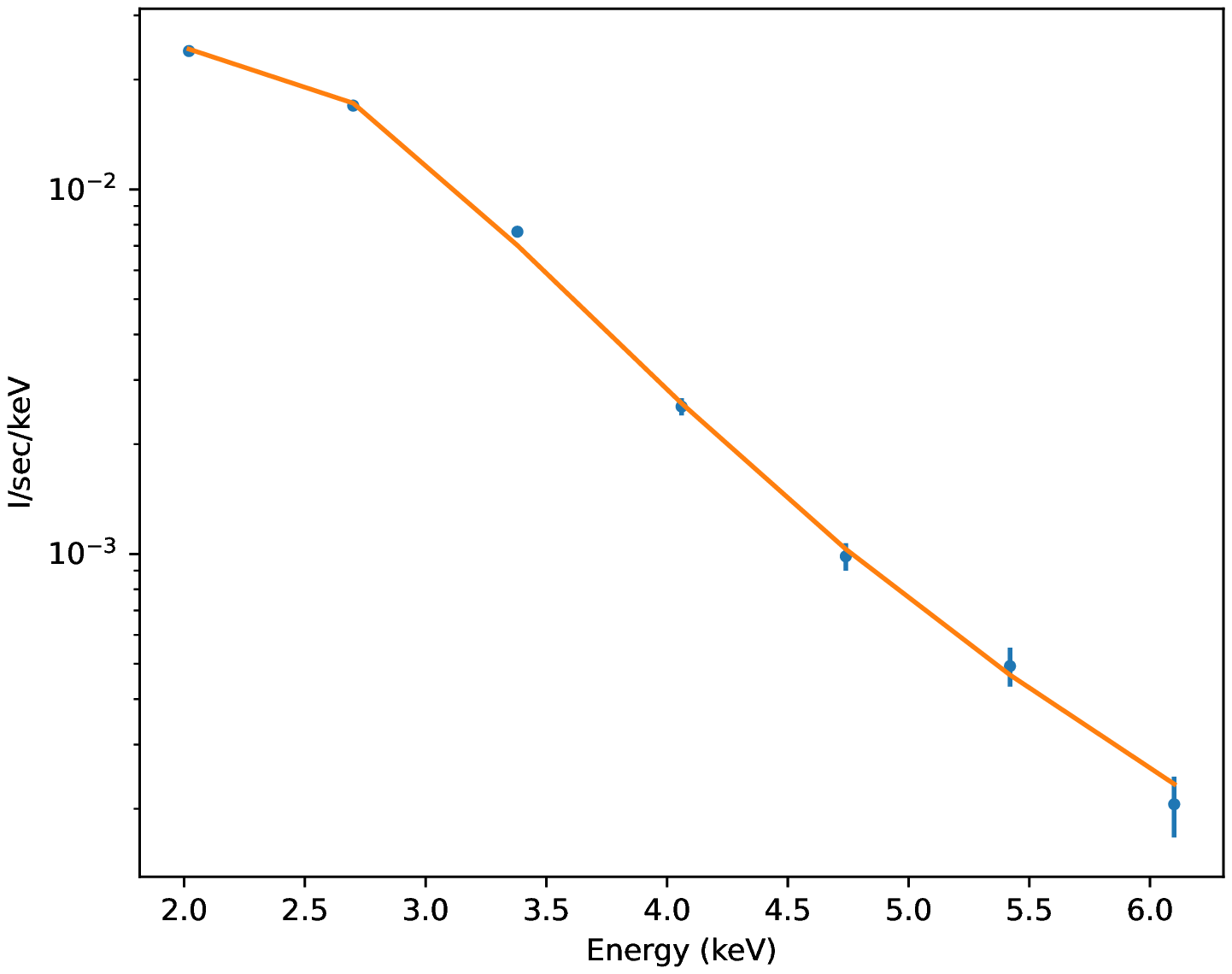}
\includegraphics[width=0.95\linewidth]{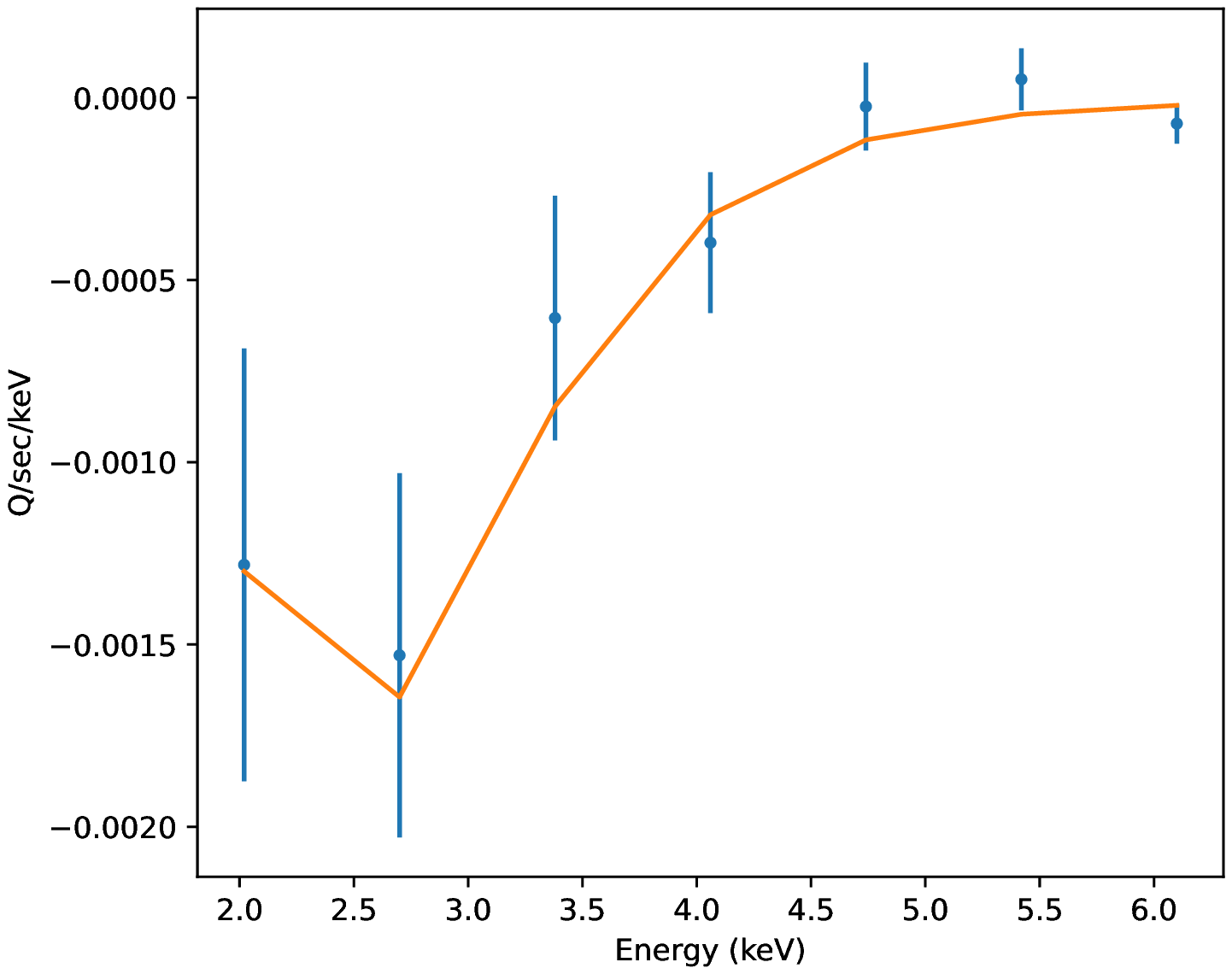}
\includegraphics[width=0.95\linewidth]{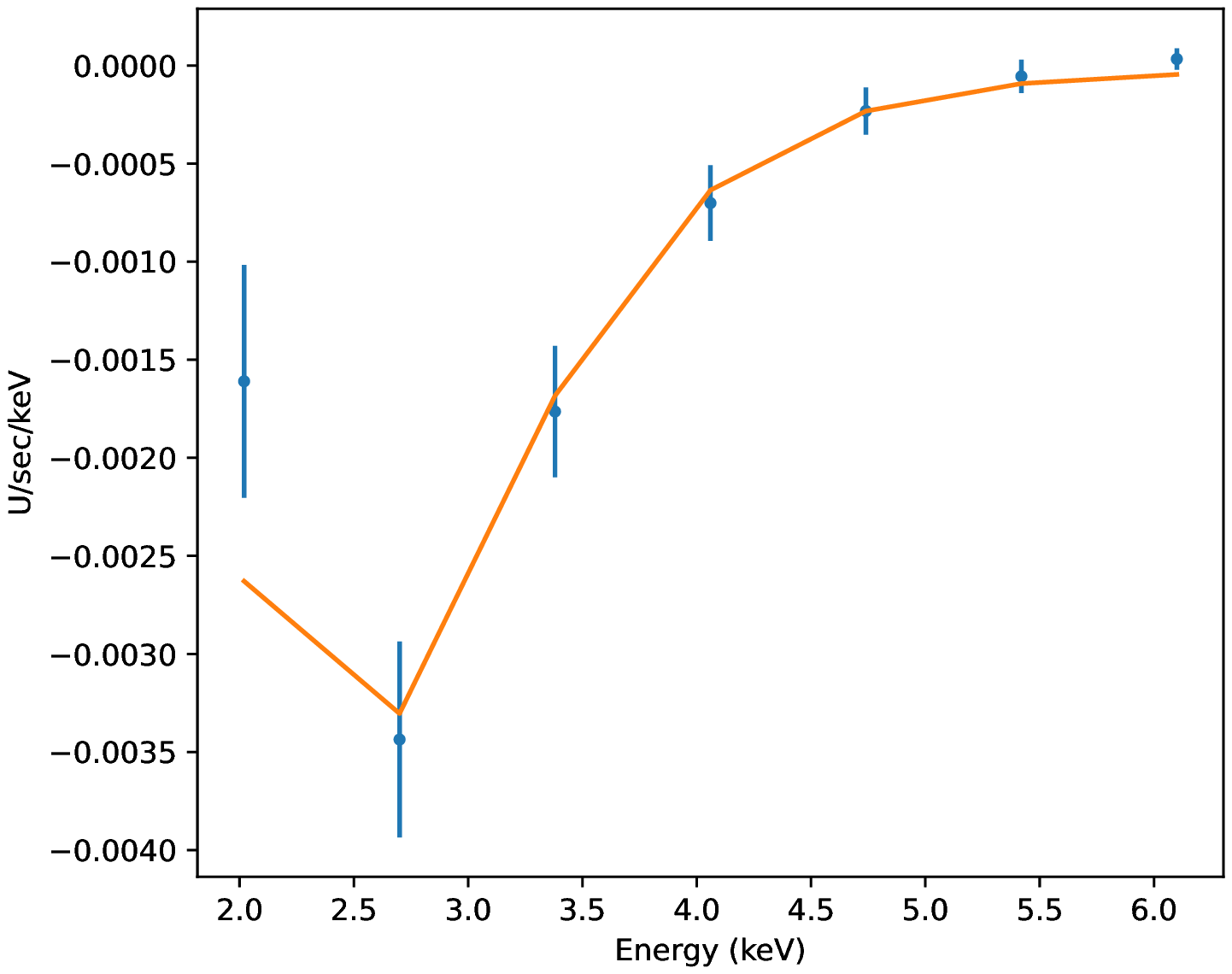}
\caption{Simulated 200\,ksec {\it IXPE} observations of AXP 1RXS J170849.0$-$400910 for the blackbody model showing the Stokes I (top), Stokes Q (center) and Stokes U (bottom) energy spectra for phase bin 3. 
The blue data points are the simulated {\it IXPE} data, and the 
orange lines are the best-fit model. For all ten phase bins, 
the fit gives a $\chi^2$ of 195.0 for 202 DoF.
\label{f:es22}}
\end{figure}
\begin{figure*}[h!]
\centering
\includegraphics[width=0.45\linewidth]{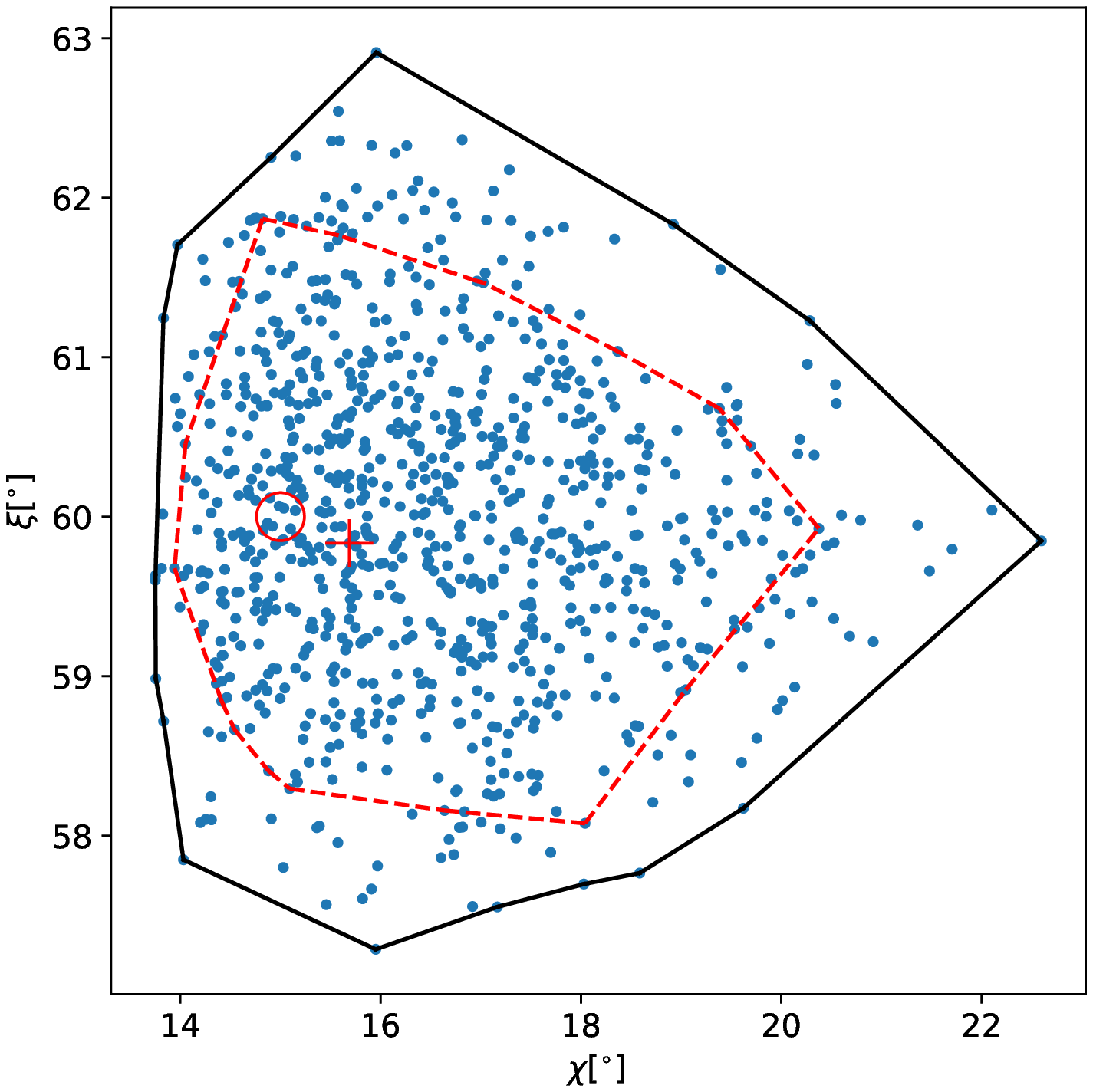}\hspace*{0.5cm}
\includegraphics[width=0.45\linewidth]{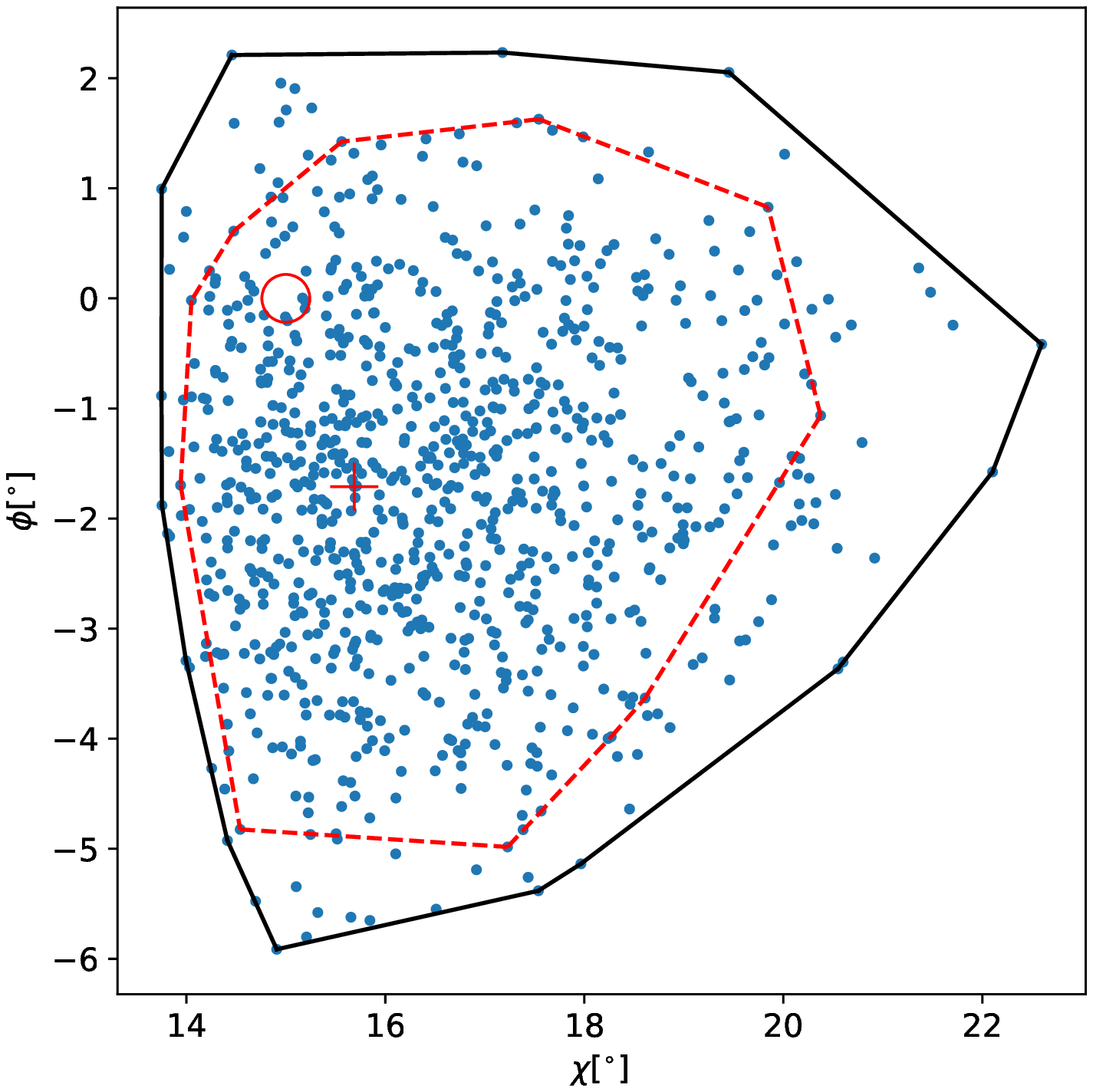}
\includegraphics[width=0.45\linewidth]{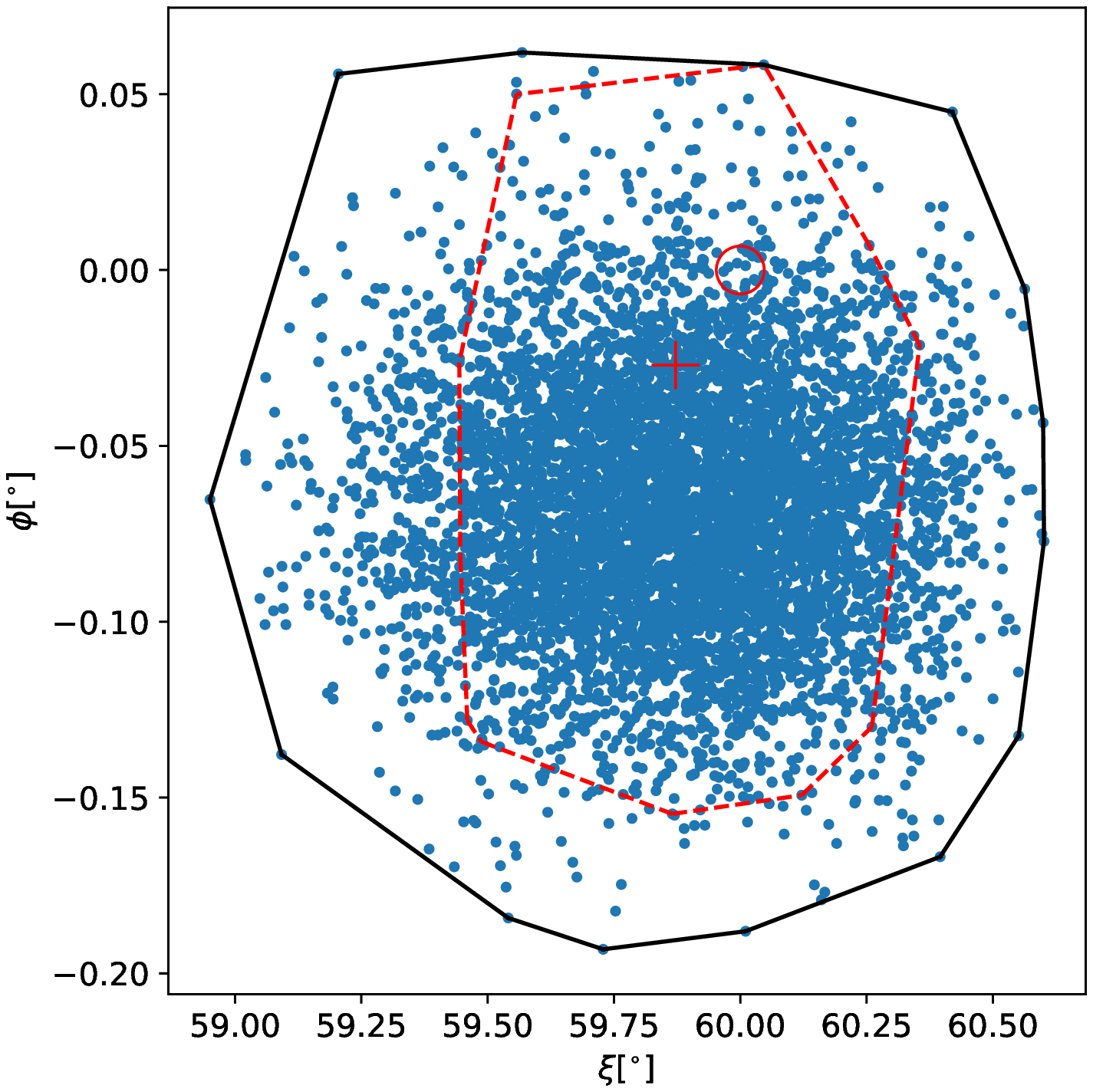}\hspace*{0.5cm}
\includegraphics[width=0.45\linewidth]{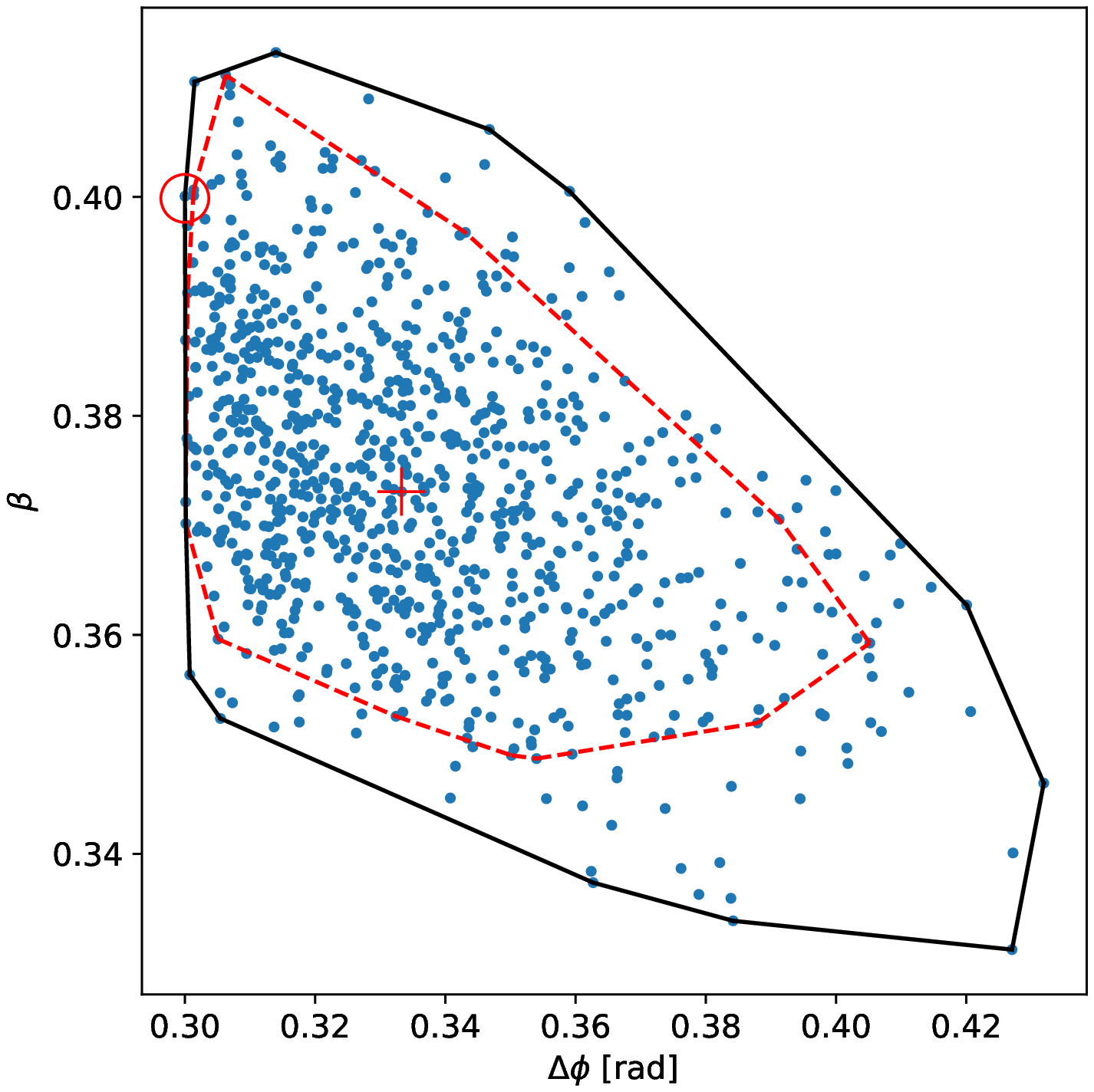}
\caption{Contour plots of the spectral parameters obtained fitting the {\it IXPE} data for the case of the blackbody model. 
The red circles show the value for which the 
data set was simulated and the red crosses show the best-fit value.
The red and black lines show the $\Delta \chi^2<13.36$ and 
$\Delta \chi^2<20.1$ contours, roughly corresponding to  
90\% and 99\% confidence intervals. 
\label{f:c22}}
\end{figure*}
\begin{table*}[h!]
\caption{Results from fitting the simulated {\it IXPE} fixed-ions condensed surface data. \label{t:cs}}
\begin{center}
\begin{tabular}{|p{5cm}|c|c|c|c|c|c|c|c|c|}
\hline
Model Name& 
$\chi^2$/DoF & 
{\tt norm} &
$\chi$  & $\xi$  & 
 $\Delta\phi$  & 
 $\beta$&
 $n_{\rm H}$ &  $\phi$ 
 &  {\tt offset}  \\
 
 & & [$10^{-9}$] & [$^{\circ}$] & [$^{\circ}$] & [rad] & [$c$] &  [$10^{22}{\rm cm}^{-2}$] & [$^{\circ}$] & 

\\ \hline
Fixed-ions cond.\,surf. (input)& NA & 8.74    & 15.0 & 60.0 & 0.50 & 0.35 & 0.513 & 0.0& 0.858\\  
Blackbody (best fit) & 435.6/202  & 12.2        & 9.8  & 58.5 & 0.73 & 0.20 & 0.212 & 16.6&0.861 \\ 
Fixed-ions cond.\,surf. (best fit) & 192.3/202   & 8.81 & 16.0 & 65.0 & 0.61 & 0.29  & 0.435 & -1.1 & 0.855 \\ 
Fixed-ions cond.\,surf. (no QED) & 193.4/202   & 8.77 & 20.5 & 62.5 & 0.54 & 0.30  & 0.504 & 14.0 & 0.856 \\  \hline

\end{tabular}
\end{center}
\end{table*}
\begin{figure}[h!]
\centering
\includegraphics[width=0.95\linewidth]{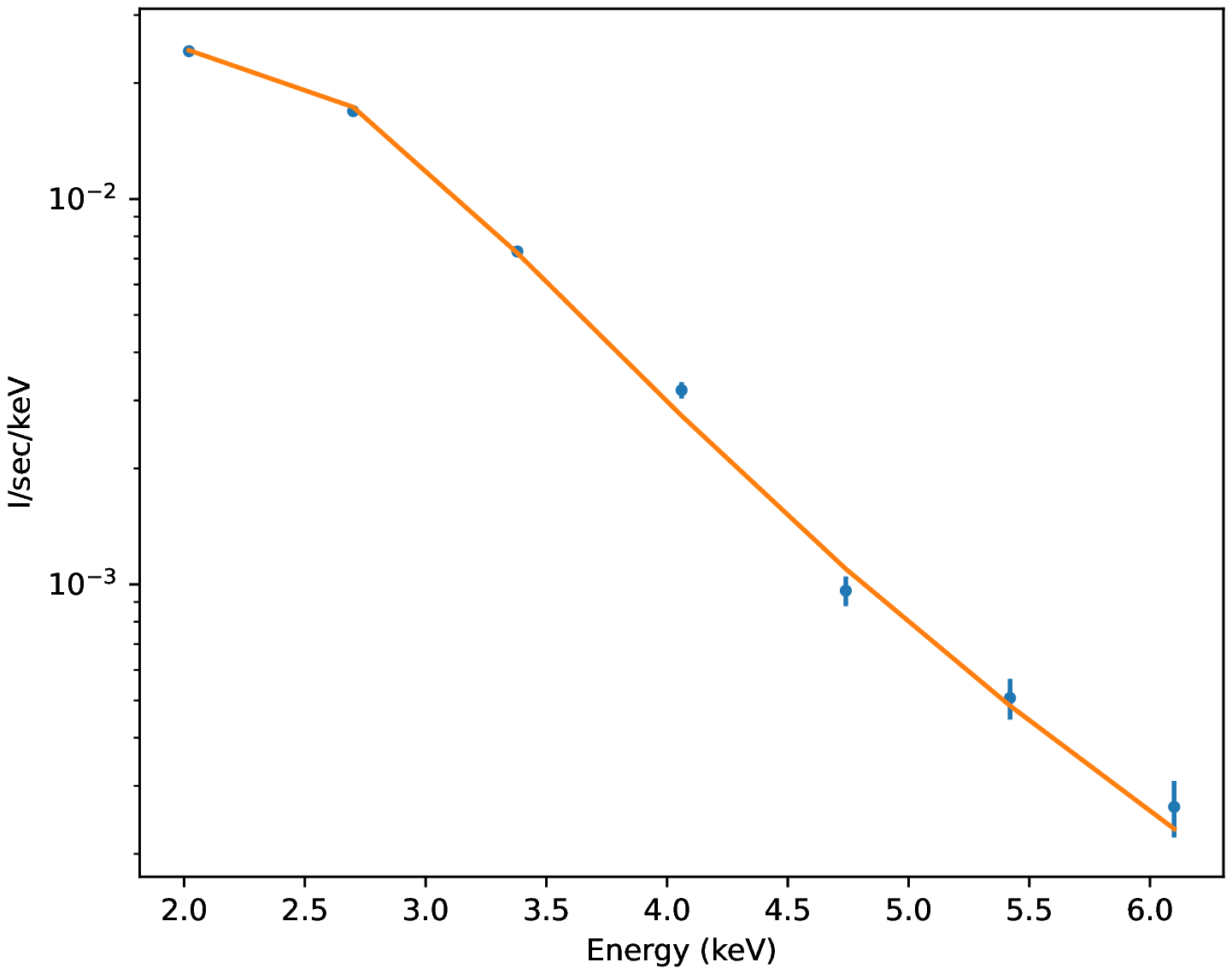}
\includegraphics[width=0.95\linewidth]{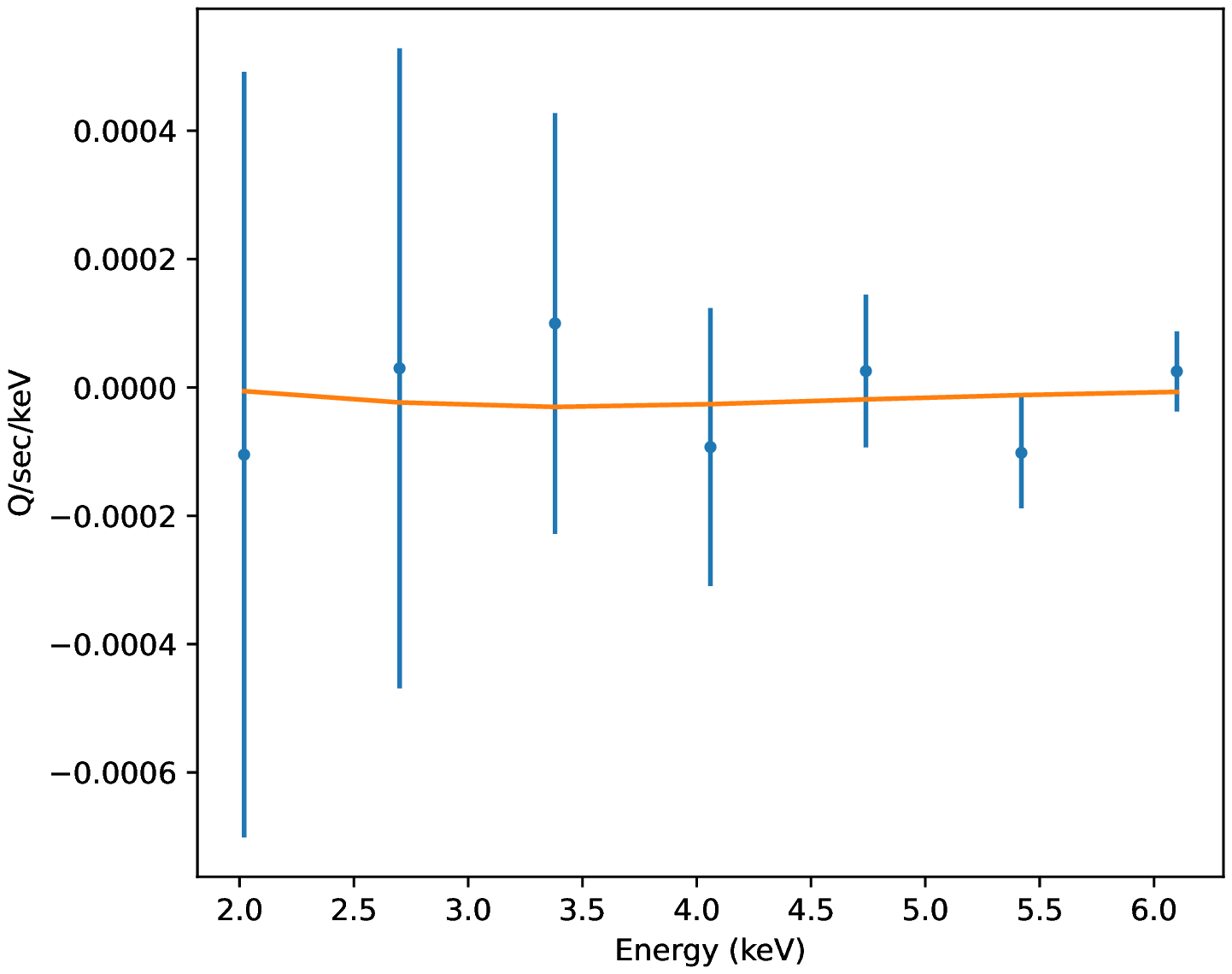}
\includegraphics[width=0.95\linewidth]{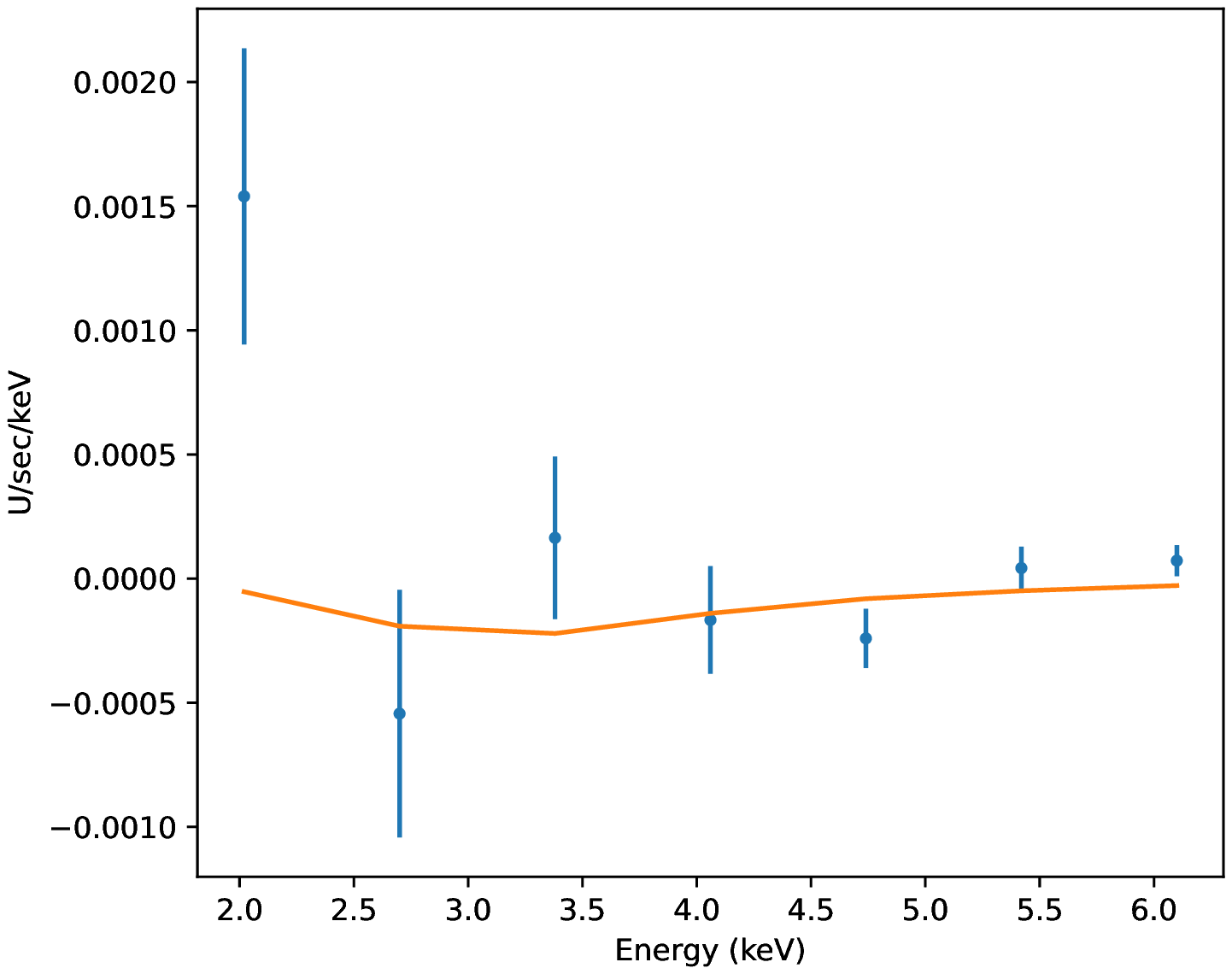}
\caption{Simulated {\it IXPE} observations of AXP 1RXS J170849.0$-$400910
assuming the best-fit fixed-ions model that fits the actual {\it XMM-Newton} data.
The blue data points show the simulated {\it IXPE} observations, and the orange
lines show the best fit model that gives a $\chi^2$ of 192.3 for 202 DoF.
\label{f:e33}}
\end{figure}
\begin{figure*}[h!]
\centering
\includegraphics[width=0.45\linewidth]{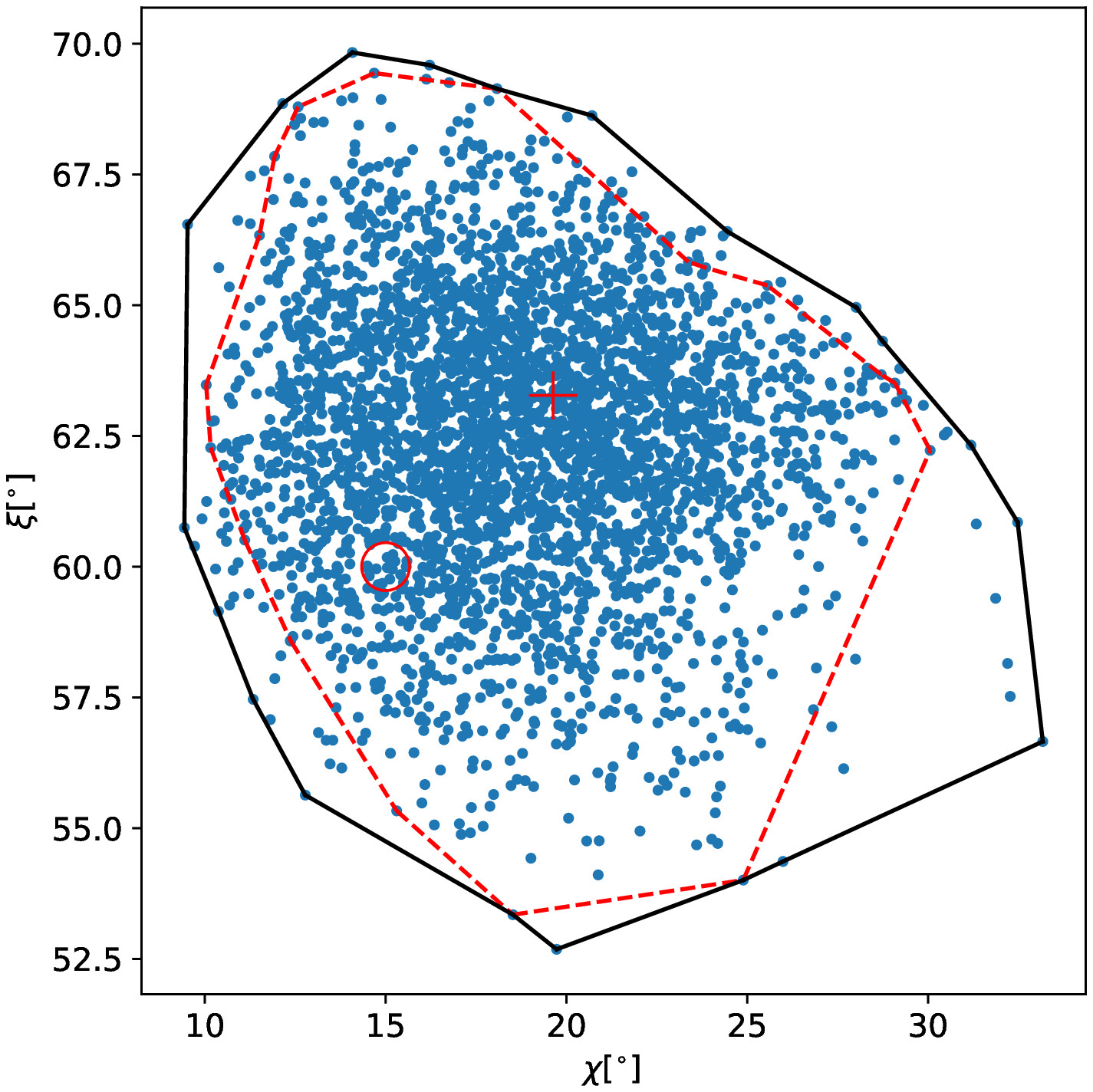}\hspace*{0.5cm}
\includegraphics[width=0.45\linewidth]{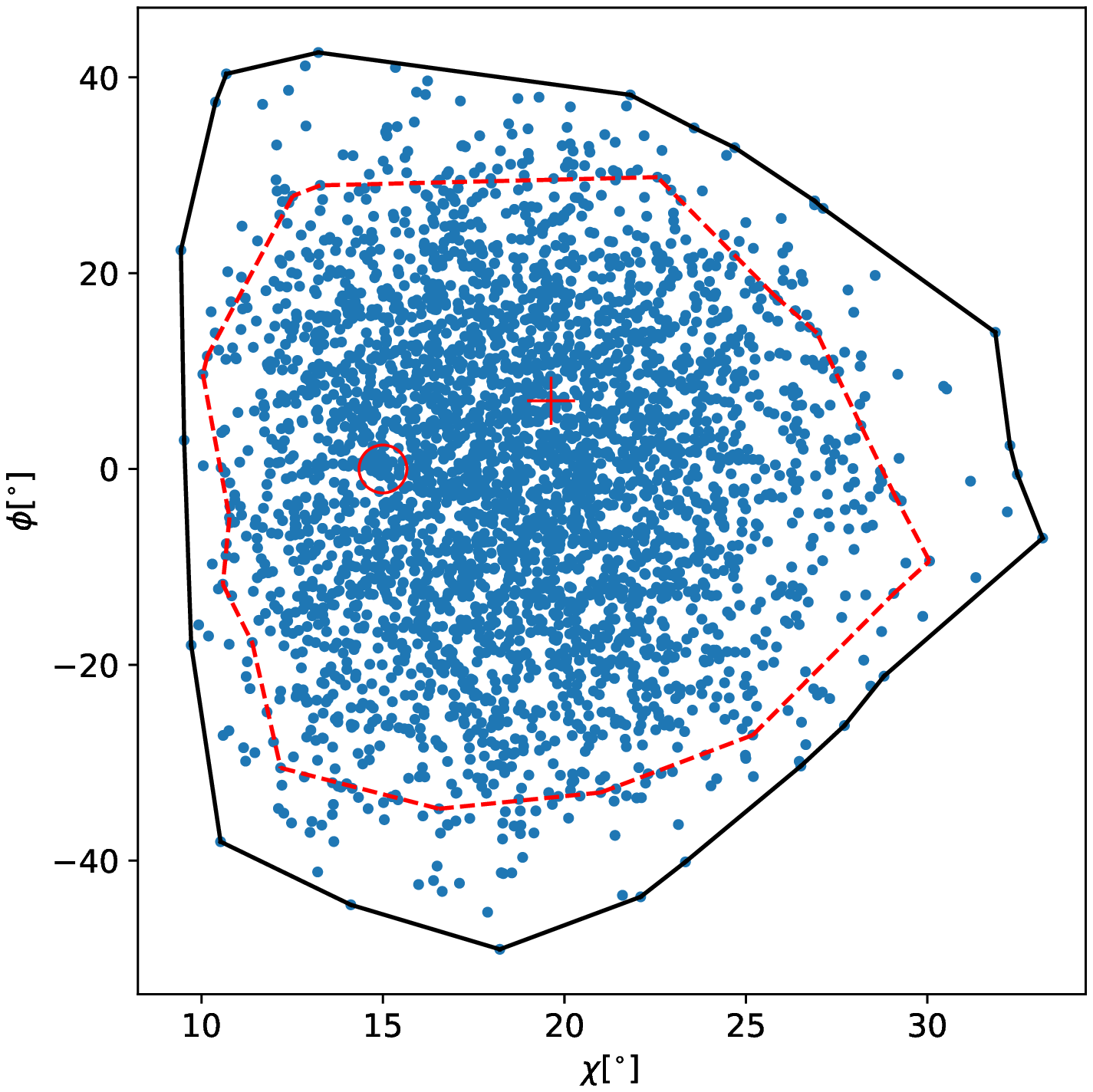}
\includegraphics[width=0.45\linewidth]{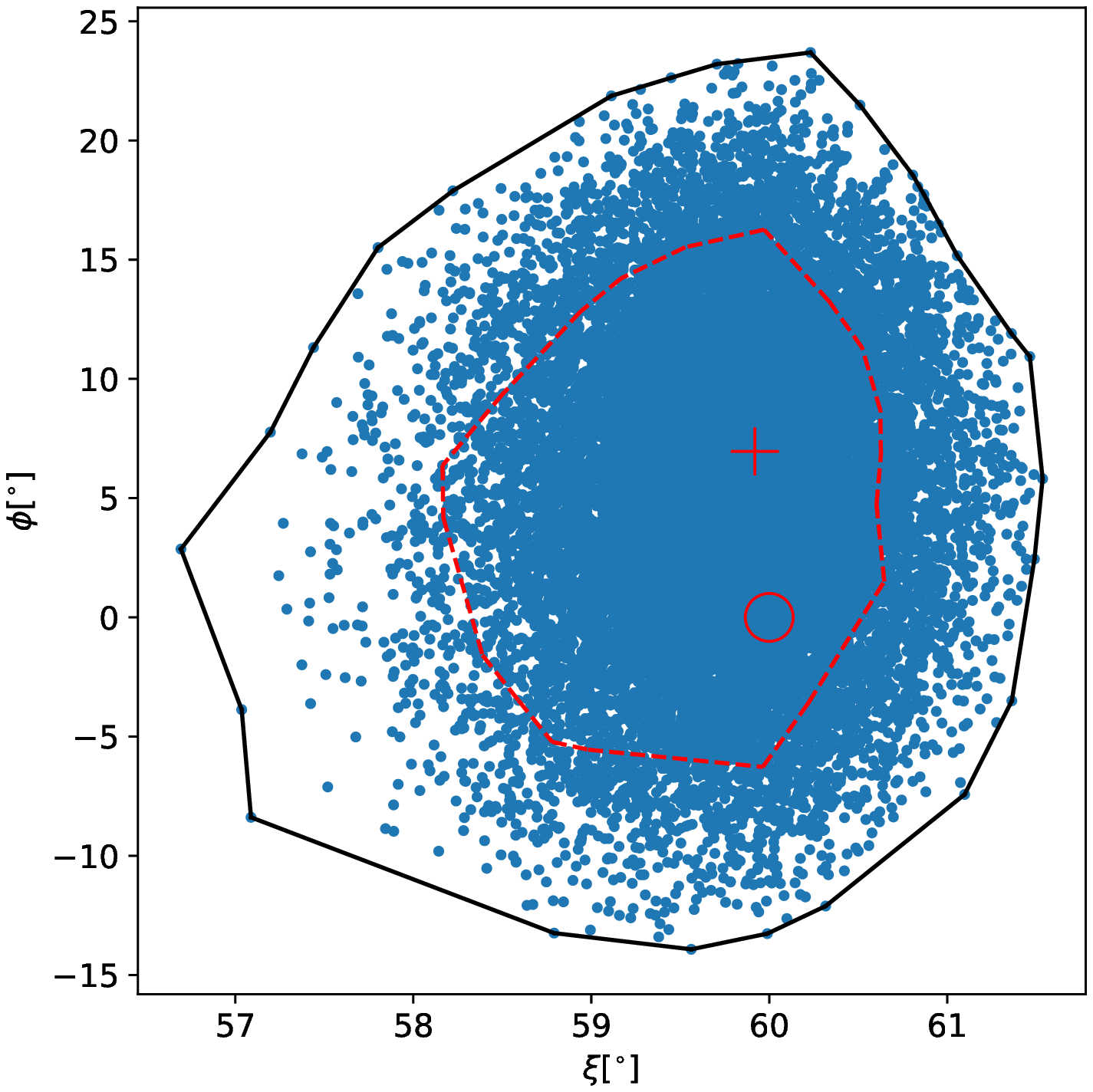}\hspace*{0.5cm}
\includegraphics[width=0.45\linewidth]{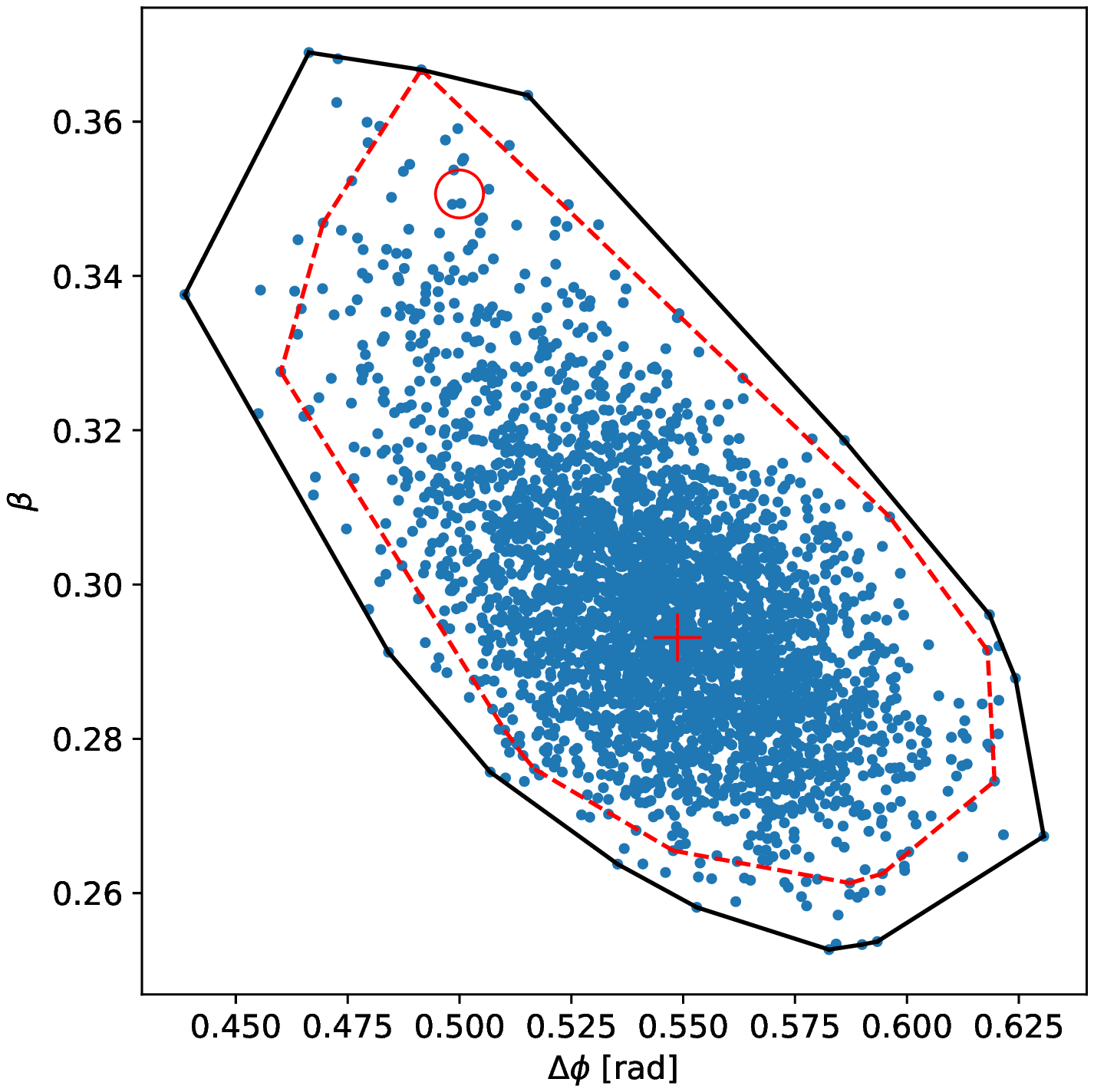}
\caption{Contour plots of the spectral parameters obtained fitting the {\it IXPE} data for the case of the fixed-ions condensed surface model. The red circles show the value for which the 
data set was simulated and the red crosses show the best-fit value.
The red and black lines show the $\Delta \chi^2<13.36$ and 
$\Delta \chi^2<20.1$ contours, roughly corresponding to  
90\% and 99\% confidence intervals. \label{f:c33}.}
\end{figure*}
For each detected or simulated event, a set of Stokes parameters
is calculated:
\begin{eqnarray}
i&=&1\\
q&=&2\cos{(2\psi)}\\
u&=&2\sin{(2\psi)}\
\end{eqnarray}
with $\psi$ being the angle of the electric field 
vector measured counterclockwise from the celestial north pole. 
The ``observed'' $I_{\rm obs}$, $Q_{\rm obs}$ and $U_{\rm obs}$ 
energy spectra are obtained by summing the Stokes parameters of the 
individual events in $N$ energy bins.
The factors of two in the expressions of $q$ and $u$ 
are chosen such that the measured values $I_{\rm obs}$, 
$Q_{\rm obs}$ and $U_{\rm obs}$ equal the Stokes $I$, $Q$, and $U$ values describing the X-ray beam \cite[][]{2015APh....68...45K}.

We developed a python code to simulate and fit {\it IXPE}
magnetar observations. The simulations use the 
Ancillary Response Function (ARF), the Redistribution Matrix File (RMF), and the Modulation Response Files (MRF)
from \citet{Baldini2020}. 
Whereas the ARF gives the effective area of the telescope as a function of the photon energy (the product of the mirror area,
blanket and other transmissivities, and the detector efficiency),
the MRF is the product of the ARF and the energy dependent
modulation factor $\mu(E)$.
The latter gives the fractional modulation of
the azimuthal scattering angle distribution for 
a 100\% polarized signal and depends on the polarimeter, 
and the event reconstruction methods.

Based on the theoretically expected $I$, $Q$ and $U$ values, 
the code simulates the observed $I_{\rm obs}$, $Q_{\rm obs}$, and $U_{\rm obs}$ values with the help of the {\it IXPE} ARF, RMF, 
and MRF, taking the $Q$ and $U$ variances and the $Q-U$ covariances
into account \cite[][]{2015APh....68...45K}.
We note that the modulation factor reduces the expected values of $Q_{\rm obs}$ 
and $U_{\rm obs}$ to $\mu$ times the Stokes $Q$ and $U$ values of the X-ray beam,
The $\chi^2$-value is calculated using $\sqrt{I}$ as the error on $I_{\rm obs}$, 
and $\sqrt{2\,I}$ as the error on the $Q_{\rm obs}$ and $U_{\rm obs}$-values \citep{2017ApJ...838...72S}.

Table \ref{t:sherpa} summarizes all the fitting parameters of the {\tt magnetar} model. In addition to the model parameters mentioned above, the model includes several additional parameters. The model predictions can be scaled with an overall 
normalization factor {\tt norm}. 
The polarization direction depends on the orientation of the
source in the sky. Accordingly, the model includes the parameter 
{$\phi$} to rotate the simulated source anticlockwise relative 
to the celestial north pole. The model allows the user to perform a phase-averaged or a phase-resolved analysis. 
In the case of the former, the user specifies the phase 
interval over which the model is averaged. 
In the case of the latter, the user specifies 
two additional parameters: a phase offset {\tt offset} 
and the phase direction {\tt dir}. 
The value {\tt offset} instructs the code to fit the data 
at phase $x$ with the model at phase $x+${\tt offset}. 
A value of {\tt dir=}+1 uses the simulated phase sequence as  modeled, and a value of {\tt dir=}-1 inverts the order of the modeled sequence 
(inverted phase $=$ 1-phase). 
The inverted phase corresponds to the emission
of a magnetar with mirror-imaged geometry.
We use the {\tt sherpa}'s {\tt levmar} and {\tt moncar}
minimization engines. 
We find that {\tt sherpa} fits the phase-resolved 
{\it XMM-Newton} data relatively quickly and reliably.
The fits of the simulated {\it IXPE} data converge 
more slowly and often end up in local minima. 
We address this issue by starting the fit from many different
initial parameter values, and choosing the fit result with the
overall lowest $\chi^2$-value.
Once the latter is found, we initiate a random exploration 
of 50,000 parameter combinations in the $\sim\pm$3$\sigma$ 
neighborhood of the best-fit parameter combination. 
This random exploration is used to map out the $\chi^2$ test statistics and thus the confidence intervals on the parameters, and to double check that the search indeed found 
the minimum of the test statistic. If the random exploration 
reveals a parameter configuration with a lower test statistics, 
the random walk is recentered on the new minimum.
\section{Results from fitting the {\it XMM-Newton} observations}\label{s:xmm}
The {\it XMM-Newton} data were acquired between 21:50:40 UTC on August 28, 2003 and 10:11:51 UTC on August 29 with a total exposure time of 44.7 ksec (obs. ID 0148690101). We use here only the data from the EPIC-pn camera, that was operated in small-window mode with the medium thickness optical filter.
The data reduction was performed using the \textsc{epproc} and \textsc{espfilt} pipelines of version 15 of the Science Analysis System (SAS), with standard parameters. We corrected the time of arrivals to the Solar System barycenter with the tool \textsc{barycen}, and then we folded them to the best fitting period of 11.00178(2) s.
The source events and the ARF and RMF were selected from a circle of radius 27'' centered on the source position, while the background was extracted from a nearby circular region of radius 40''. We obtained the spectra corresponding to 10 phase bins, and we rebinned them using the \textsc{grppha} tool with a minimum of 30 counts per bin.

The results from fitting the {\it XMM-Newton} data with the 
phase and angle averaged magnetar models are shown in Table \ref{t:xmmA} and Fig.\  \ref{f:XMMspecA}. The blackbody, free-ions condensed surface, and fixed-ions condensed surface models give the best fits with $\chi^2/$Degrees of Freedom (DOF)-values of
1123/67, 2373/67 and 859/67, respectively. 
The magnetized atmosphere models gives a much poorer fit with  
a $\chi^2/$DoF-values of 39239/67. 
The poor fit of the magnetized atmosphere model is a recurring 
finding of the analyses shown in this paper. 
The magnetized atmosphere model predicts a thermal 
component that is too broad.

The results from fitting the phase-resolved {\it XMM-Newton} data with the 
phase and angle resolved magnetar model are shown in Table \ref{t:xmm}.
The fixed-ions condensed surface model gives the best fit ($\chi^2$/degree of freedom (DoF) $=$ 3103.9/703) followed by the blackbody model ($\chi^2$/DoF  $=$ 3372.5/703)
and the free-ions condensed surface model ($\chi^2$/DoF $=$ 3728.9/703). Compared with
the other ones, the magnetized atmosphere gives by far the poorest fit, with a
$\chi^2$/DoF-values of 37432.5/703.
For all models, the fit strongly prefers the inverted pulse sequence ({\tt dir}$=$-1), 
indicating that the geometry of AXP 1RXS J170849.0$-$400910 is the mirror image 
of the one assumed in our model. As this result is consistent across all considered
models, we do not mention it any more in the following.
For the blackbody model, we repeated the fit with the best-fit $\chi$ and $\xi$ 
values of the fixed ion condensed surface model. 
The $\chi^2$/DoF increases from 3372.5/703 to 3419.4/703 showing that within the 
rather large systematic uncertainties (and thus very large $\chi^2$-values), 
a wide range of angles $\chi$ and $\xi$ gives rather similar $\chi^2$-results.

Figure \ref{f:XMMspec} presents the energy spectra of a particular phase bin (i.e., phase bin 3 out of 10), showing that the 
free-ions condensed surface model (lower left panel), the
fixed-ions condensed surface model 
(lower right panel) and the blackbody model (upper left panel) 
give the best 
description of the spectral shape. The magnetized atmosphere model (upper right panel) fits the data poorly.
All of the models give large $\chi^2/$DoF-values, and thus reveal systematic
differences between the model and the observations that are larger than the
statistical errors of the {\it XMM-Newton} EPIC-pn data. We discuss potential reasons for these systematic shortcomings of the models in the discussion section.
Figure \ref{f:lc1} compares the observed and modeled pulse profiles. The shapes of the observed and modeled lightcurves agree rather well.
The overall normalization is off for the magnetized atmosphere model,
owing to the different impact of different energy ranges on the lightcurve 
and the spectral fits.

Figure \ref{f:all3} shows the best fixed-ions condensed surface fit for all ten phase bins.
For phase bins 1 and 10, we observe strong deviations in the 1--2 keV energy range. The model underpredicts the thermal emission.
For phase bins 4 and 5, the model undepredicts the powerlaw emission.
Finally, for phase bins 7 and 8, the model overpredicts the 2--4 keV emission. 
We also tried to add a power law or a broken power law component to the fits. 
We report here only the results for
the fixed ion condensed surface model. Adding a power law to each of the ten {\it XMM-Newton} phase bins, adds 20 new fitting parameters. The fit improves from a $\chi^2$/DoF$=$ $3103.9/703=$ 4.42 (without power laws) to $\chi^2/$DoF$=$ $2308.8/683=$ 3.38 (with power laws). Adding broken power laws, the fit improves further to $\chi^2/$DoF$=$ $1829.7/673=$ 2.72.
Before adding the (broken) power law models, the main discrepancies come from the 
magnetar models predicting too soft energy spectra for some of the phase bins 
(Fig. \ref{f:XMMspec}).  After adding the (broken) power law models, the main discrepancies come from the other phase bins for which the model predicts too hard energy spectra.

Fitting the phase-averaged magnetar models to the phase-averaged {\it XMM-Newton} data also gives rather poor results.
For example, the fit with the fixed-ions condensed surface plus broken power law model gives $\chi^2$/DoF$=$ 506.1$/$71$=$8.03 for $\chi=0.1^{\circ}$, $\xi=45^{\circ}$, $\Delta\phi=0.32$, $\beta=0.43$, $n_{\rm H}=0.554 \times 10^{22}$\,atoms cm$^{-2}$, $\Gamma_1=-10$ (frozen) and $\Gamma_2=2.34$.
Similarly, for the blackbody model, we get $\chi^2$/DoF$=$ 561.8$/$71$=$8.92 for $\chi=0.1^{\circ}$, $\xi=30.4^{\circ}$, $\Delta\phi=0.3$, $\beta=0.48$, $n_{\rm H}=0.614 \times 10^{22}$\,atoms cm$^{-2}$, 
$\Gamma_1=-10$ (frozen) and $\Gamma_2=1.51$.

We can summarize the findings from this section as follows:
the fixed-ions condensed surface and the blackbody models succeed to approximately describe the observed energy spectra and light curves. However, the {\it XMM-Newton} data have such a good signal-to-noise ratio, that the $\chi^2$-values are entirely systematics limited, and the models do not give a statistically valid description of the data. Although the addition of power laws and broken power laws improve the fits
significantly, they are still not acceptable from a statistical point of view. As the fits are systematics 
limited, we refrain from deriving confidence intervals for the parameters.   
\section{Analysis of simulated {\it IXPE} observations}\label{s:ixpe}
We use the best-fit parameter values determined from fitting the {\it XMM-Newton} data
and the magnetar emission models to generate simulated {\it IXPE} data sets.
All our simulations assume an {\it IXPE} integration time of 200\,ksec.
The reader should keep in mind that {\it IXPE} can acquire even deeper observations.
Using the parameters from the {\it XMM-Newton} fits implies that we assume the same flux 
level as observed in August 2003. This is a reasonable assumption because the source
shows only little variability and restricted to the hard X-ray range \cite[][]{gotz+07,sho+14}.

Figure \ref{f:polMod} shows the predicted Stokes $I$, $Q$ and $U$ 
energy spectra and the corresponding polarization fraction energy spectra
for all four magnetar emission models. We show the results for only 
one phase bin (phase bin 3). The blackbody and magnetized 
atmosphere models (the upper two panels of Fig.\ \ref{f:polMod})  are characterized by  very high polarization fractions between 60\% and 100\% 
from 2\,keV to 5\, keV, that is over much of {\it IXPE's} energy range. 
The polarization fractions of the two condensed surface models 
(two lower panels of Fig.\ \ref{f:polMod}) are well below 10\% between $\sim2$--$3$\,keV 
and then rise to values exceeding 10\% at higher energies.
In the following, we focus on simulations
for the two best-fitting models: the blackbody model and the fixed-ion
condensed surface model.

\subsection{Fitting the simulated blackbody {\it IXPE} data} 

Table \ref{t:bb} summarizes the fitting results of a simulated blackbody model.
The blackbody model gives a good fit with a $\chi^2$-value of 195.0 for 202 DoF. Figure \ref{f:es22} presents the simulated and modeled Stokes $I$, $Q$ and $U$ distributions (phase bin 3).
The plots allow the reader to gauge the magnitude of the statistical 
errors of the {\it IXPE} data.
Figure \ref{f:c22} shows how well we can determine the model
parameters. For 8 model parameters of interest, 
$\Delta \chi^2$-values of $<13.36$ and $<20.1$ give rough estimates of the
90\% and 99\% confidence intervals \citep{1976ApJ...210..642A}.
Here and in the following we give accuracies on 99\% confidence level.
We infer that the two angles $\chi$ and $\xi$ are determined to accuracies of
$\pm$4$^{\circ}$ and $\pm$3$^{\circ}$, respectively. 
The orientation of the source in the sky can be determined to an accuracy 
of $\pm$4$^{\circ}$. 
The fits constrain $\Delta\phi$ and $\beta$ 
to accuracies of $\pm$0.06\,rad and 0.035 $c$, respectively.
The results for $\Delta\phi$ and $\beta$ are somewhat correlated.

The simulations show that if the emission follows the blackbody model, the 
fixed-ions condensed surface model can clearly be excluded.
The $\chi^2$/DoF increases from 195.0/202 for the blackbody model to 
754.6/202 for the fixed-ions condensed surface model.
The Stokes $I$, $Q$ and $U$ energy spectra show that 
the deviations are significant in all three Stokes parameters. 

\subsection{Fitting the simulated fixed-ions condensed surface {\it IXPE} data} 

Table \ref{t:cs} summarizes the results for the simulated fixed-ions condensed
surface model. The fixed-ions condensed surface model gives an acceptable fit
with a $\chi^2$ of 192.3 for 202 DoF (Fig.\  \ref{f:e33}). 
In this case, the blackbody model can be excluded at a high degree of confidence, as it
gives a $\chi^2$ of 435.6 for 202 DoF.
For the particular model, Fig.\  \ref{f:c33} shows the parameter regions for which
the $\chi^2$-values deviate by less than 13.36 ($\approx$1\,$\sigma$) and
20.1 ($\approx$90\% confidence level) from the minimum.
The parameters $\chi$, $\xi$, $\phi$, $\Delta\phi$, and $\beta$ can be constrained 
with 90\% confidence intervals of 
$\pm11^{\circ}$, 
$\pm8^{\circ}$,
$\pm45^{\circ}$,
$\pm$0.09, and
$\pm$0.06 $c$, respectively.
Again, we see a correlation between $\Delta\phi$ and $\beta$. 

\subsection{Impact of QED effects}
We fit the blackbody and the fixed-ions condensed surface 
{\it IXPE} observations simulated including QED effects with the models 
obtained with and without the QED effects.
The results are also given in Tables \ref{t:bb} and \ref{t:cs}.
For the simulated blackbody observations, the blackbody model with QED fits
significantly better ($\chi^2/$DoF$=$195.0/202) than the model without QED 
($\chi^2/$DoF$=$711.3/202). 

The situation is different for the simulated fixed-ions condensed surface {\it IXPE}
observations: fitting the fixed-ions condensed surface model with QED effects gives
us a $\chi^2/$DoF$=$192.3/202 and the model without QED effects gives
a $\chi^2/$DoF$=$193.4/202. Interestingly, although  
the $\chi^2$ does not change much, the best-fit parameters do change, indicating that
the almost identical $\chi^2$-value is somewhat of a chance coincidence.
The results show that {\it IXPE} would have good chances to detect the presence of QED effects if the blackbody model applies. For the fixed-ions condensed surface, the 
overall low polarization fractions make it more challenging to do so, requiring longer integration times.

\section{Conclusions}\label{s:disc}
In this paper, we present four different magnetar models, and use them to fit the {\it XMM-Newton} observations of the AXP 1RXS J170849.0$-$400910. 
Although the models give a good qualitative fit of the observations, 
the fits are not satisfactory from a statistical point of view.
Furthermore, several of our fits end up at the edges of the simulated
parameter ranges. Additional power law components lead to smaller $\chi^2$-values,
but systematic discrepancies remain.
The results strongly prefer the fixed-ions condensed surface model 
and the blackbody model over the magnetized atmosphere and free-ions condensed surface models. 
The energy spectra predicted by the magnetized atmosphere model 
show much broader peaks than the observed ones, and seems to be ruled out.  

The remaining discrepancies between the best-fit model and the observations indicate
that the model does not capture completely
all the aspects of the system. 
We remark that our model predictions rely on a series of simplifying
approximations. First of all, we considered the NS surface temperature as
a constant. The fact that the low-energy part of the observed phase-averaged 
spectrum cannot be adequately reconstructed by all the models (see Fig.\  
\ref{f:XMMspecA} and Table \ref{t:xmmA}) may indeed indicate that the real
surface temperature distribution is more complicated. Further approximations
concern the modeling of the stellar magnetosphere. Our simulations assume that
the motion of the magnetospheric electrons is described by a unidirectional
flow, composed by a single-temperature Maxwellian distribution superimposed
to a 1-dimensional bulk motion. Both assumptions affect the slope of the
powerlaw tail in the high-energy part of the spectrum, so that they may
need to be relaxed to achieve more satisfactory fits. We also note that
our model assumes a globally-twisted, axisymmetric external field, and a
more complicated magnetic field topology may be required to reproduce the
real situation.
We would like to emphasize that the {\it XMM-Newton} data set has an excellent signal to noise ratio.

We present predictions for the polarization that the upcoming {\it IXPE}
observatory will measure. Whereas the blackbody and magnetized atmosphere models 
predict very high ($\sim 100\%$) polarization fractions over the 
entire {\it IXPE} energy range, the condensed surface model predicts polarization fractions well below 10\% below 4\,keV, rising to values 
exceeding 10\% only above 4\,keV. 
The simulations show that the {\it IXPE} observations will allow us to cleanly decide between the 
high-polarization (magnetized atmosphere and blackbody) 
and low-polarization (condensed surface) models.
Our studies show that QED effects are clearly detectable for the high-polarization
models. The QED effects could be detected 
for the low-polarization models with longer exposures than considered here.   
\begin{acknowledgements}
      The authors thank Luca Baldini for making and sharing the
      {\it IXPE} ARF, MRF, and RMF files, and the {\it IXPE} team
      for highly enjoyable discussions and meetings. 
      The work would not have been possible without
      the {\tt CIAO} and {\tt sherpa} software packages and the
      excellent documentation provided by the authors.
      HK acknowledges support by NASA through the grants NNX16AC42G and 80NSSC18K0264. Part of this work was supported by the German \emph{Deut\-sche For\-schungs\-ge\-mein\-schaft, DFG\/} project number Ts~17/2--1. The work of RT, RT, SM and MR is partially supported by the Italian MUR through grant UNIAM (PRIN 2017LJ39LM). 
      
\end{acknowledgements}

%
%

\end{document}